\documentclass[aps,prb,twocolumn,showpacs,preprintnumbers,superscriptaddress]{revtex4}
\usepackage{amssymb,amsfonts,euscript,times,mathptm}
\usepackage[intlimits]{amsmath}
\usepackage{graphicx}
\usepackage{pstricks,pst-3d}
\usepackage{psfrag}
\usepackage{calc}

\newcommand{\intd}[1]{\!\!\! #1 \,\,\,\,}
\renewcommand \vec \pmb

\begin{document}

\title{Polar magneto-optical Kerr effect for low-symmetric ferromagnets}
\author{Helmut Rathgen}
\email{helmut.rathgen@web.de}
\affiliation{Institut f\"ur Mathematische Physik, Technische Universit\"at
  Braunschweig, Mendelssohnstra{\ss}e 3, D--38106 Braunschweig, Germany}
\affiliation{Condensed Matter Theory, Department of Physics, Uppsala University, Box 530,
  S--75121 Uppsala, Sweden}
\author{Mikhail I. Katsnelson}
\affiliation{Condensed Matter Theory, Department of Physics, Uppsala University, Box 530,
  S--75121 Uppsala, Sweden}
\affiliation{Institute for Molecules and
Materials, Radboud University Nijmegen,
Toernooiveld 1, 6525 ED, The Netherlands}
\author{Olle Eriksson}
\affiliation{Condensed Matter Theory, Department of Physics, Uppsala University, Box 530,
  S--75121 Uppsala, Sweden}
\author{Gertrud Zwicknagl}
\affiliation{Institut f\"ur Mathematische Physik, Technische Universit\"at
  Braunschweig, Mendelssohnstra{\ss}e 3, D--38106 Braunschweig, Germany}

\begin{abstract}
The polar magneto-optical Kerr effect (MOKE) for low-symmetric
ferromagnetic crystals is investigated theoretically based on
first-principle calculations of optical conductivities and a transfer
matrix approach for the electrodynamics part of the problem. Exact
average magneto-optical properties of polycrystals are described, taking
into account realistic models for the distribution of domain
orientations. It is shown that for low-symmetric ferromagnetic single
crystals the MOKE is determined by an interplay of crystallographic
birefringence and magnetic effects. Calculations for single and
bi-crystal of hcp $\langle 11\bar 20\rangle$ Co and for a polycrystal
of CrO$_2$ are performed, with results being in good agreement with
experimental data.
\end{abstract}


\maketitle

\section{Introduction}

The magneto-optical Kerr effect (MOKE)
is a versatile method to probe magnetic properties of thin
films. Advanced by the rapid developments in crystallographic
growth techniques, a variety of low-symmetric crystalline surfaces
have been subject to MOKE measurements in the last decades. This
has led to systematic investigations of magneto-optical anisotropy
effects \cite{weller94prl72:2097}.

State of the art theoretical approaches to investigate the MOKE are based on
first-principle calculations of dielectric tensors in the framework of
the Kubo-Greenwood formalism
\cite{kubo57jpsj12:570,greenwood58ppsl71:585} as suggested by Wang and
Callaway \cite{wang74prb9:4897}. The MOKE is obtained from a
dielectric tensor by means of an approximative analytic expression
\begin{equation}\label{common_moke}
  \psi+i\chi
  =
  \frac{\epsilon_{xy}}{(1-\epsilon_{xx})\sqrt{\epsilon_{xx}}}.
\end{equation}
derived originally by Argyres in 1955 \cite{argyres55pr97:334}. $\psi$
denotes the Kerr rotation and $\chi$ denotes the Kerr ellipticity.

This approach requires in general that the dielectric tensor has
symmetry
\begin{equation}\label{common_symmetry}
  \epsilon=
  \begin{pmatrix}
    \epsilon_{xx} & \epsilon_{xy} & 0 \\
    -\epsilon_{xy} & \epsilon_{xx} & 0 \\
    0 & 0 & \epsilon_{zz} \\
  \end{pmatrix}.
\end{equation}
There have been theoretical attempts to extend the approach to
low-symmetric systems, however so far the complete electrodynamics
calculation for low symmetric dielectric tensors has not been
considered.

There are many interesting ferromagnets that have a low
symmetry, e.g. CrO$_2$, hcp $\langle 11\bar 20\rangle$ Co and FePt
grown in the $\langle 010\rangle$ direction. All of these systems have
two different crystallographic axis in the surface plane, so beside
their magneto-optical activity they exhibit crystallographic
birefringence.

In this paper we show that for such crystals it is important to
consider the complete optical response including birefringence and
magnetic effect in order to describe correctly the polar MOKE.
Further, we show that the optical response is qualitatively different for
single- and
polycrystals and finally, for polycrystals it
sensitively depends on the ordering of
crystallographic domains. We calculate the MOKE of hcp
$\langle 11\bar 20\rangle$ Co and of $\langle 010\rangle$
CrO$_2$. For Co we show that the previous interpretation of experimental
data of anisotropic polar MOKE \cite{weller94prl72:2097} in
terms of a manifestation of magneto-crystalline anisotropy remains valid.

The paper is organised as follows. In the subsequent section we
describe our approach to the complete calculation of the
electrodynamics problem by means of transfer matrix
methods. Theoretical description of ellipsometry measurements for
single- and polycrystals is given
in Sec.~\ref{sec:ellipsometry}. In Sec.~\ref{sec:first_principles}
we discuss first-principle calculations of optical conductivities.
Space-time symmetry of Co and CrO$_2$ crystals is described in
Sec.~\ref{sec:symmetry}. The calculated optical response of Co and CrO$_2$
is presented in Sec.~\ref{sec:co} and Sec.~\ref{sec:cro2} respectively.
In Sec.~\ref{sec:conclusions} a summary and conclusions are given.

\section{Transfer matrix methods} \label{electrodynamics}

The optical response of a finite system of layers to an incident
plane wave can be described by transfer matrix methods
\cite{born80principles_of_optics,zak90jap68:4203,stepanov00prb61:15302}.
The description is valid if the magnetic permeability is unity and the
wavelength of the light is large compared to the microscopic structure
of materials and also large compared to interface roughness.
In the most general case a system with $n$ boundaries is described by a
regular set of $4n$ linear equations that determines the complex
amplitude vectors of all plane waves in all media. We
briefly describe the method.

We first choose a coordinate system such that the $z$-axis is the
surface normal and the scattering plane is spanned by the $z$-axis and
the $y$-axis. In the half space of the incident and reflected wave
Fresnel's secular equation reads
\begin{equation}
  -\vec k^2\vec E + \frac{\omega^2}{c^2}\epsilon\vec E = \vec 0.
\end{equation}
We substitute $\vec r\mapsto\frac{\omega}{c}\vec r$ and define
$q:=\frac{c}{\omega}k_y$ and $k:=\frac{c}{\omega}k_z$. This gives
\begin{equation}
\begin{aligned}
  q &= \sqrt{\epsilon}\,\sin\vartheta\\
  k &= \pm \sqrt{\epsilon}\,\cos\vartheta=:\pm k_0,
\end{aligned}
\end{equation}
where $\vartheta$ is the incident angle.
This gives an ansatz for the wave
\begin{equation}\label{ansatz_scalar}
  \vec E=\vec E^{in}e^{i(qy-k_0z-\omega t)} + \vec
  E^{refl}e^{i(qy+k_0z-\omega t)},
\end{equation}
where $\vec E^{in}$ is the known amplitude vector of the incident wave
and the complex amplitude vector of the reflected wave satisfies
\begin{equation}\label{e_k_scalar}
  E^{refl}_z=-\frac{q}{k_0}E^{refl}_y,
\end{equation}
leaving two free parameters $E^{refl}_x$ and $E^{refl}_y$.
For other media the most general plane
wave solution to Maxwell's equations is a combination of four
independent waves.
In the case of a scalar medium it is
\begin{equation}
  \vec E=\vec E^{1}e^{i(qy+k^1z-\omega t)} + \vec
  E^{2}e^{i(qy+k^2z-\omega t)},
\end{equation}
where
\begin{equation}\label{k_scalar}
  k^{1,2}=\pm\sqrt{\epsilon-q^2}
\end{equation}
and the $x$- and $y$-components of $\vec E^1$ and $\vec E^2$ are independent.
In the case of a tensor medium it is
\begin{equation}\label{ansatz_tensor_general}
  \vec E = a^1\vec n^1e^{i(qy+k^1z-\omega t)}\,+\,\dots\,
    +\, a^4\vec n^4e^{i(qy+k^4z-\omega t)}
\end{equation}
with four free parameters $a^1,\dots,a^4$ satisfying
\begin{equation}\label{a_n}
  \vec E^{i}=a^{i}\vec n^{i}.
\end{equation}
$k^1,\dots , k^4$ are the roots of the fourth order polynomial in $k$
\begin{equation}\label{e_k_tensor_matrix}
  \textnormal{Det}\begin{vmatrix}
    \epsilon_{xx}-q^2-k^2 & \epsilon_{xy} & \epsilon_{xz} \\
    \epsilon_{yx} & \epsilon_{yy}-k^2
      & \epsilon_{yz} + q\,k \\
    \epsilon_{zx} & \epsilon_{zy}+q\,k & \epsilon_{zz}-q^2\\
  \end{vmatrix}
  = 0
\end{equation}
and the vectors $\vec n^1,\dots ,\vec n^4$ are associated kernels.

In the half space on the backside of the layers two waves can always be
discarded. For transparent medium these are two backward travelling
waves, for an absorbing medium these are two exponentially decaying waves.

In our case (a bulk metallic system with no intermediate layer) we
have only an absorbing tensor
half space and the ansatz for the waves in the responding system reduces to
\begin{equation}\label{ansatz_tensor}
  \vec E= a^1\vec n^1e^{i(qy+k^1z-\omega t)} + a^2\vec
    n^2e^{i(qy+k^2z-\omega t)},
\end{equation}
where $k^1$ and $k^2$ are the roots that have negative imaginary parts
(negative $z$-direction corresponds to forward travelling waves).

Stressing the assumption of unity magnetic permeability, four
independent boundary conditions follow from Maxwell's equations
stating that
\begin{equation}\label{boundary_conditions}
  E_x,\hspace{1ex}E_y,\hspace{1ex}\partial_zE_x
  \textnormal{\hspace{1em}and\hspace{1em}}iqE_z-\partial_zE_y
\end{equation}
are continuous.

Substituting the ansatz,
Eq.~(\ref{ansatz_scalar}) and Eq.~(\ref{ansatz_tensor}) in the
boundary conditions, we get
\begin{widetext}
\begin{equation}\label{linear}
  \begin{pmatrix}
    -1 & 0 & n^1_x & n^2_x \\
     0 & -1 & n^1_y & n^2_y \\
    -k_0 & 0 & k^1n^1_x & k^2n^2_x\\
     0 & \frac{q^2}{k_0}+k_0 & qn^1_z-k^1n^1_y & qn^2_z-k^2n^2_y
  \end{pmatrix}
  \begin{pmatrix}
    E^{refl}_x\\
    E^{refl}_y\\
    a^1\\
    a^2
  \end{pmatrix}
  =
  \begin{pmatrix}
    E^{in}_x\\
    E^{in}_y\\
    -k_0E^{in}_x\\
    qE^{in}_z+k_0E^{in}_y
  \end{pmatrix}.
\end{equation}
\end{widetext}
This is a regular system of four linear equations. Stressing
Eq.~(\ref{e_k_scalar}) and Eq.~(\ref{a_n}) its solution
determines the complex amplitudes vectors of all waves.

We have written a numerical implementation of the most
general case of a transfer matrix approach (based on standard
LAPACK\cite{anderson99lapack_users_giude} routines
and polynomial solver\cite{press86numerical_recipes}). It is described
in detail in Ref.~\onlinecite{rathgen03diplomarbeit}.

\section{Ellipsometry for single- and polycrystals}\label{sec:ellipsometry}

The state of polarization of a plane wave is conveniently
described by Stokes parameters \cite{born80principles_of_optics}
\begin{equation}\label{stokes_parameters}
  \vec S
  =
  \begin{pmatrix}
    S_0\\
    S_1\\
    S_2\\
    S_3\\
  \end{pmatrix}
  =
  \begin{pmatrix}
    E_x\overline{E}_x+\overline{E}_yE_y\\
    E_x\overline{E}_x-\overline{E}_yE_y\\
    E_x\overline{E}_y+\overline{E}_xE_y\\
    i(E_x\overline{E}_y-\overline{E}_xE_y)\\
  \end{pmatrix},
\end{equation}
where \mbox{$\vec E=(E_x,E_y)$} is the complex amplitude vector of the
plane wave in
the coordinate system of the polarization state analysis.

The state of polarization of a set of incoherent plane waves that add by
their intensities is described by the sum of their Stokes parameters.
Both for a single wave and for an incoherent wave, the rotation angle of the
polarization ellipse $\psi$ and
its ellipticity $\chi$ are related to the Stokes parameters by
\begin{equation}\label{rotation}
  \tan2\psi=\frac{S_2}{S_1}
\end{equation}
and
\begin{equation}\label{ellipticity}
  \sin2\chi=\frac{S_3}{\sqrt{S_1^2+S_2^2+S_3^2}}.
\end{equation}
In the general case the polarization ellipse is the intensity behind
an analyser for all positions. Only in the special case of a single
wave is this equivalent to the curve that is drawn by the tip of
the electric field vector.

The optical response of a polycrystal can be described by the sum over
Stokes parameters of single crystalline domains weighted by surface
areas of the domains and intensities shining on
them\cite{uspenskii96prb54:474}. The sum extends
over all domains that are illuminated in the experiment. The approach
is valid if
single crystalline domains are large compared to the wavelength.
We can calculate Stokes parameters for polycrystals by summing over
Stokes parameters obtained from transfer matrix calculations for
single crystals.

\section{First-principle calculations of optical conductivities} \label{sec:first_principles}

We briefly describe the calculation of optical constants by means of
first-principle calculations. Our approach is basically standard
unless we evaluate the Kubo--Greenwood formula directly without
Kramers--Kronig transformation and analytical continuation (see also
Ref.~\onlinecite{rathgen04psT109:170}).

In this section we consider the optical conductivity \mbox{tensor
  $\sigma$} rather than the corresponding dielectric tensor
$\epsilon$. The quantities are related by the identity
\begin{equation}
  \epsilon_{\alpha\beta}(\omega)=
  \delta_{\alpha\beta}+i\tfrac{4\pi}{\omega}\sigma_{\alpha\beta}(\omega).
\end{equation}

In general, intra-band, as well as direct and indirect inter-band
transitions, contribute to the optical conductivity. Spins may
flip (for magnetic dipole transitions) or stay constant (for
electric dipole transitions ) during excitations. It is a common
practice to account only for the contribution of electric dipole
(non-spin-flip) direct inter-band transitions by means of {\em ab
initio} methods while treating the contribution of intra-band
transitions by a phenomenological Drude term
\begin{equation}
  \sigma_{D}(\omega)=\frac{\sigma_0}{1+\omega^2\tau^2},
\end{equation}
and neglecting all other contributions
\cite{wang74prb9:4897,oppeneer92prb45:10924,gasche96prb53:296,blaha01wien2k}.
A broad variety of linear optical and magneto-optical effects in
metals as well as in semiconductors have been successfully
described in the framework of this approximation, see e.g.
Refs.~\onlinecite{ebert96rpp59:1665,kunes02prb65:165105} and
references therein. In the transition metals, intra-band
transitions turn out to be important in the range from 0 eV up to
0.5 eV \cite{maksimov88jpf18:833}. It is shown in
Ref.~\onlinecite{wang74prb9:4897} that a corresponding Drude
contribution is negligible for energies larger than 1 eV in
the case of Ni. Throughout this work we neglect any
phenomenological Drude contribution.

The Kubo--Greenwood expression for the contribution of direct inter band transitions
to the optical conductivity reads \cite{oppeneer92prb45:10924}
\begin{widetext}
\begin{equation}\label{kubo}
\sigma_{\alpha\beta}(\omega)=\frac{ie^2}{m^2\hbar}\int_{BZ}\intd{d^3k}\hspace{-1em}
  \sum_{\substack{l\,,\,n\\E_l(\vec k)<E_F\\E_n(\vec k)>E_F}}\hspace{-1em}
  \frac{1}{\omega_{nl}(\vec k)}\left[    \frac{\Pi_{ln}^{\alpha}(\vec k)\Pi_{nl}^{\beta}(\vec k)}
                                                         {\omega-\omega_{nl}(\vec k)+\frac{i}{\tau(\omega)}} +
  \frac{ \left( \Pi_{ln}^{\alpha}(\vec k)\Pi_{nl}^{\beta}(\vec k) \right)^*}
                 {\omega+\omega_{nl}(\vec k)+\frac{i}{\tau(\omega)}}      \right],
\end{equation}
\end{widetext}
where the indices $l$ and $n$ denote the spin and all band quantum
numbers for the occupied and empty states respectively and $\vec
k$ is the quasi momentum running through the Brillouin zone, $E_F$ is the
Fermi energy. The symbol $\Pi_{nl}^{\alpha}(\vec
k)\,,\,\,\alpha=x, y, z$ denotes the matrix elements of the
momentum operator given below by Eq.~(\ref{matrixelements}), and
$\omega_{nl}(\vec k)$ is the energy difference between the
involved states,
\begin{equation}
  \omega_{nl}(\vec k)=\frac{1}{\hbar}\left( E_{n}(\vec k) - E_{l}(\vec k)
  \right).
\end{equation}
Finally, $\tau(\omega)$ is a phenomenological relaxation time.
Throughout this work we use a constant relaxation time of 0.136
eV. The results of this paper are insensitive to the actual choice of
this value.

Together with the energy differences $\omega_{nl}(\vec k)$, the
matrix elements of the momentum operator are obtained from the
underlying band structure calculation by evaluating the expression
\begin{equation}\label{matrixelements}
  \vec \Pi_{ln}(\vec k)=\int\intd{d^3r}\psi_l^*(\vec k,\vec r)\left[ \vec p + \frac{\hbar}{4mc^2}\left[ \vec\sigma \times
     \vec\nabla V(\vec r) \right]\right]\psi_n(\vec k,\vec r)
\end{equation}
Here $\psi_n(\vec k,\vec r)$ is the Bloch wave function with
quantum numbers as described above, $\vec p = -i\hbar\nabla$ and
$V(\vec r)$ is a crystal potential. State-of-art works on {\em ab
initio} calculated optical constants neglect the spin--orbit term
in the expression for the matrix elements of the momentum
operator, Eq.~(\ref{matrixelements}). This has been found to be a
good approximation, see, e.g., Ref.~\onlinecite{wang74prb9:4897}.
We follow this approach.

Expression~(\ref{kubo}) may be computed directly or via symmetrized
limit expressions requiring Kramers--Kronig transformations and
analytical continuation to finite relaxation times. We recently
discussed advantages and
disadvantages of both approaches that become important when the
conductivity tensor has low symmetry
\cite{rathgen04psT109:170}. In the present paper Expression~(\ref{kubo}) is
computed directly.

For electronic structure calculation we use a relativistic
full-potential linear muffin-tin orbital (FP-LMTO) code. The code is
described in detail in
Ref.~\onlinecite{wills00dreysse148}. A discussion of the treatment
of spin--orbit coupling by means of the second variational step
can be found in Ref.~\onlinecite{koelling77jpc10:3107}.

\section{Symmetry considerations} \label{sec:symmetry}

\begin{figure}
\begin{center}
\rput(-3.5cm,-3.5cm){\textbf a)}
\rput(-3.5cm,-8.5cm){\textbf b)}
\rput(3.1cm,-.5cm){$\underline m m\underline 2$}
\rput(3.3cm,-5.5cm){$\underline 2/\underline m$}
\begin{pspicture}(-1.7cm,-1.9cm)(2.1cm,1.9cm)


\psset{unit=3.5cm}

\rput{91}(.5,0){
\scalebox{.5}{
\newcommand{\vpoint}{0.1 -1 0.3}
\psset{viewpoint=\vpoint}
\psset{normal=0 0 1}
\psset{linewidth=.02}
  \ThreeDput{
    \pspolygon(-.5,-.866)(-1,0)(-.5,.866)(.5,.866)(1,0)(.5,-.866)
    \psline[linewidth=.005,linestyle=dashed](-.5,-.866)(.5,.866)
    \psline[linewidth=.005,linestyle=dashed](-1,0)(1,0)
    \psline[linewidth=.005,linestyle=dashed](-.5,.866)(.5,-.866)
  }
  \ThreeDput(0,0,1.633){
    \pspolygon(-.5,-.866)(-1,0)(-.5,.866)(.5,.866)(1,0)(.5,-.866)
    \psline[linewidth=.005,linestyle=dashed](-.5,-.866)(.5,.866)
    \psline[linewidth=.005,linestyle=dashed](-1,0)(1,0)
    \psline[linewidth=.005,linestyle=dashed](-.5,.866)(.5,-.866)
  }
\psset{arrowscale=1.5}
  \ThreeDput[normal=0 -1 0](0,0,0){
    \psline{->}(0,0)(0,.5)
  }
  \ThreeDput{
    \psline{->}(0,0)(0,-1.2)
  }
  \ThreeDput[normal=0 -1 0]{
    \psline{->}(0,0)(.6,0)
  }
  \ThreeDput[normal=\vpoint](0.1,0.2,.47){\rput{-90}{\scalebox{2}{$x$}}}
  \ThreeDput[normal=\vpoint](-0.1,-1.2,-.1){\rput{-90}{\scalebox{2}{$y$}}}
  \ThreeDput[normal=\vpoint](.64,0,-0.07){\rput{-90}{\scalebox{2}{$z$}}}
  \ThreeDput(0,0,1.633){
    \psline(-.5,-.866)(.5,-.866)
  }
  \ThreeDput[normal=0 -1 0](0,0,0){
    \psline(-1,0)(-1,1.633)
    \psline(1,0)(1,1.633)
  }
  \ThreeDput[normal=0 -1 0](0,.866,0){
    \psline(-.5,0)(-.5,1.633)
    \psline(.5,0)(.5,1.633)
  }
  \ThreeDput[normal=0 -1 0](0,0,.8165){
    \psline{->}(-.2,0)(.2,0)
  }
  \ThreeDput[normal=\vpoint](.25,-.3,1.1){\rput{-90}{\scalebox{1.5}{$\vec M$}}}
  \ThreeDput[normal=0 -1 0](0,-.866,0){
    \psline(-.5,0)(-.5,1.633)
    \psline(.5,0)(.5,1.633)
  }
\psset{dotsize=.1}
\psset{normal=\vpoint}
  \ThreeDput(-.5,-.866,0){
    \psdots(0,0)
  }
  \ThreeDput(-.5,-.866,1.633){
    \psdots(0,0)
  }
  \ThreeDput(.5,-.866,0){
    \psdots(0,0)
  }
  \ThreeDput(.5,-.866,1.633){
    \psdots(0,0)
  }
  \ThreeDput(-.5,.866,0){
    \psdots(0,0)
  }
  \ThreeDput(-.5,.866,1.633){
    \psdots(0,0)
  }
  \ThreeDput(.5,.866,0){
    \psdots(0,0)
  }
  \ThreeDput(.5,.866,1.633){
    \psdots(0,0)
  }
  \ThreeDput(-1,0,0){
    \psdots(0,0)
  }
  \ThreeDput(-1,0,1.633){
    \psdots(0,0)
  }
  \ThreeDput(1,0,0){
    \psdots(0,0)
  }
  \ThreeDput(1,0,1.633){
    \psdots(0,0)
  }
  \ThreeDput(0,0,0){
    \psdots(0,0)
  }
  \ThreeDput(0,0,1.633){
    \psdots(0,0)
  }
  \ThreeDput[normal=0 -1 0](-.5,.2887,0){
    \psline[linewidth=.005,linestyle=dashed](0,0)(0,.8165)
  }
  \ThreeDput[normal=0 -1 0](.5,.2887,0){
    \psline[linewidth=.005,linestyle=dashed](0,0)(0,.8165)
  }
  \ThreeDput[normal=0 -1 0](0,-.5773,0){
    \psline[linewidth=.005,linestyle=dashed](0,0)(0,.8165)
  }
  \ThreeDput(-.5,.2887,.8165){
    \psdots(0,0)
  }
  \ThreeDput(.5,.2887,.8165){
    \psdots(0,0)
  }
  \ThreeDput(0,-.5773,.8165){
    \psdots(0,0)
  }
}
}
\end{pspicture}
\begin{pspicture}(-3cm,-3cm)(3cm,3cm)
  \psset{unit=.8cm}
  \rput(0,0){\includegraphics[scale=.28,trim= 2.1cm 5cm 2cm 8cm
      ,clip]{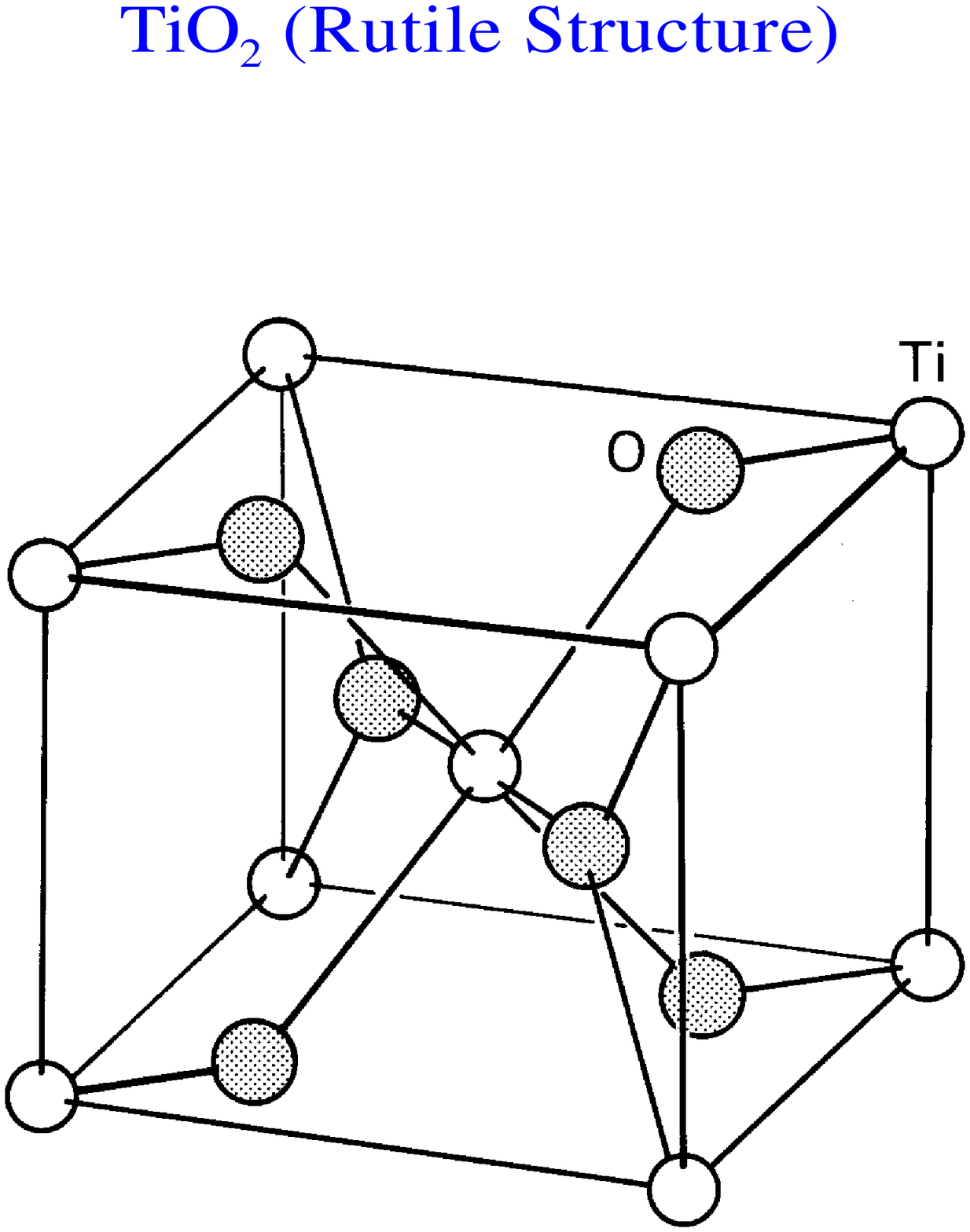}}
  \psline{-}(0,.74)\psline{->}(0,.83)(0,1.4)
  \psline{->}(-1,-1)
  \rput{-7}(0,0){
    \psline{-}(.73,0)
    \psline{-}(.83,0)(1.075,0)
    \psline{->}(1.18,0)(1.4,0)
  }
  \rput(-.2,1.55){$x$}
  \rput(-.75,-.5){$y$}
  \rput(1.4,-.4){$z$}
  \rput{-7}(0,0){\psline[linewidth=.07]{->}(1.9,0)(2.9,0)}
  \rput(3,0){$\vec M$}
  \rput(2,-2){$a$}
  \rput(-.8,-2.4){$a$}
  \rput(-2.8,-.4){$c$}
  \psdot[dotsize=.35,linecolor=white](2.5,2.3)
  \psdot[dotsize=.35,linecolor=white](.78,1.75)
  \rput(2.5,2.3){Cr}
  \rput(.78,1.8){O}
\end{pspicture}
\caption{Crystal structures of {\textbf a)} $\langle 11\bar 20 \rangle$
  ferromagnetic hcp Co and {\textbf b)} $\langle 010 \rangle$ ferromagnetic CrO$_2$.
  Coordinate systems are as used in magneto-optics calculation.}
\label{fig:structure}
\end{center}
\end{figure}

We have used standard space-time symmetry analysis
\cite{birss63rpp26:307} to find the irreducible forms of the
dielectric tensors of hcp Co with magnetisation along $\langle 11\bar
20 \rangle$ and of CrO$_2$ with magnetisation along $\langle
010\rangle$. The crystal structures are shown in
Fig.~\ref{fig:structure}. We
find space-time point groups $\underline m m
\underline 2$ and $\underline 2/\underline m$ for Co and CrO$_2$
respectively. Making the coordinate systems explicit,
irreducible sets of point group operations can be chosen as
{\em identity}, {\em $2$-fold rotation
around $z$ followed by space inversion} and {\em $2$-fold rotation
around $y$ followed by time inversion} for Co and {\em
identity}, {\em space
  inversion} and {\em 2-fold rotation around x followed by time
  inversion} for CrO$_2$. Standard symbols are $1$,
$\overline{2}z$, $\underline{2}y$ and 1, $\bar 1$, $\underline 2 x$
respectively.

Irreducible space-time symmetries of the
dielectric tensors follow by Neumann's principle which states that
\begin{equation}\label{neumann}
  \epsilon=\sigma\circ\epsilon\circ\sigma^{-1}
\end{equation}
has to be satisfied for any symmetry operator $\sigma$. For classical
point group operators the respective matrix equation can be evaluated.
For non-classical operators $\sigma=s\circ\tau$ composed of a classical
operator $s$ and the
time inversion operator $\tau$, Eq.~\ref{neumann} can be brought in
matrix form by stressing the equivalence of time inversion
and magnetisation reversal,
\begin{equation}
  \tau\circ\epsilon(\vec M)\circ\tau^{-1}=\epsilon(-\vec M)
\end{equation}
and Onsager's relation,
\begin{equation}
  \epsilon(-\vec M)=\epsilon^T(\vec M),
\end{equation}
where $\phantom{\epsilon}^T$ denotes the transpose.

For the Co crystal we find
\begin{equation}\label{symmetry_full_co}
  \epsilon =
  \begin{pmatrix}
    \epsilon_{xx}  &  \epsilon_{xy} &  0            \\
    -\epsilon_{xy} &  \epsilon_{yy} &  0            \\
    0              &  0             &  \epsilon_{zz}
  \end{pmatrix}.
\end{equation}
For CrO$_2$ we have
\begin{equation}
  \epsilon =
  \begin{pmatrix}
     \epsilon_{xx} &  \epsilon_{xy} &  \epsilon_{xz}  \\
    -\epsilon_{xy} &  \epsilon_{yy} &  \epsilon_{yz}  \\
    -\epsilon_{xz} &  \epsilon_{yz} &  \epsilon_{zz}
  \end{pmatrix}.
\end{equation}

Next we consider the symmetry properties of the same crystals but
without magnetism. Co has the well known point group $6/mmm$ and
irreducible form of the dielectric tensor without magnetism is
\begin{equation}
  \epsilon =
  \begin{pmatrix}
    \epsilon_{xx}  &  0             &  0            \\
    0              &  \epsilon_{yy} &  0            \\
    0              &  0             &  \epsilon_{yy}
  \end{pmatrix}.
\end{equation}
The CrO$_2$ crystal without magnetism is {\em non-symmomorphic}. It has
space group $P4_2/mmm$. Evaluation of Neumann's principle is standard
for pure point group operators. For symmetry operators
$\hat\sigma=\sigma\circ T$ that are a
combination of a point group operator $\sigma$ and the translation
operator $T$ (the $4$-fold screw axis $4x\circ T(c/2,0,0)$ in our
case) Neumann's principle can be evaluated by stressing the invariance
of the dielectric tensor under arbitrary translations.
\begin{equation}
  T\circ\epsilon\circ T^{-1}=\epsilon.
\end{equation}
We find the irreducible form of the dielectric tensor without
magnetism is the same as for Co.

Next we consider the expansion of the dielectric tensor in powers of
the magnetisation and stress the following symmetry properties: The zero
order contribution has symmetry of the non-magnetic
crystal. Magnetic contributions of odd order have space-time symmetry
of the magnetic crystal and are anti symmetric. Magnetic contributions
of even order have space-time symmetry of
the magnetic crystal and are symmetric. Anti symmetry respectively
symmetry property of odd and even order magnetic contributions are
arrived at in general by applying Onsager's relation to the expansion.

We find that up to second order in the magnetisation the expansion has
the symmetry, for Co,
\begin{equation}\label{symmetry_expansion_co}
  \epsilon
  =
  \begin{pmatrix}
    \epsilon_{xx}^0 & 0 & 0 \\
    0 & \epsilon_{yy}^0 & 0 \\
    0 & 0 & \epsilon_{yy}^0 \\
  \end{pmatrix}
  +
  \begin{pmatrix}
    0 & \epsilon_{xy}^1 & 0 \\
    -\epsilon_{xy}^1 & 0 & 0 \\
    0 & 0 & 0 \\
  \end{pmatrix}
  +
  \begin{pmatrix}
    \epsilon_{xx}^2 & 0 & 0 \\
    0 & \epsilon_{yy}^2 & 0 \\
    0 & 0 & \epsilon_{zz}^2 \\
  \end{pmatrix}
\end{equation}
and for CrO$_2$
\begin{equation}
  \epsilon
  =
  \begin{pmatrix}
    \epsilon_{xx}^0 & 0 & 0 \\
    0 & \epsilon_{yy}^0 & 0 \\
    0 & 0 & \epsilon_{yy}^0 \\
  \end{pmatrix}
  +
  \begin{pmatrix}
    0 & \epsilon_{xy}^1 & \epsilon_{xz}^1 \\
    -\epsilon_{xy}^1 & 0 & 0 \\
    -\epsilon_{xz}^1 & 0 & 0 \\
  \end{pmatrix}
  +
  \begin{pmatrix}
    \epsilon_{xx}^2 & 0 & 0 \\
    0 & \epsilon_{yy}^2 & \epsilon_{yz}^2 \\
    0 & \epsilon_{yz}^2 & \epsilon_{zz}^2 \\
  \end{pmatrix}.
\end{equation}
Results of
standard electronic structure
calculations are for both systems tensors of the form
\cite{guo94prb50:R10377,oppeneer95jmmm148:298,uspenskii96prb54:474} (see also
Secs.~\ref{sec:co_conductivity},\ref{sec:cro2_conductivity})
\begin{equation}\label{symmetry_calculation}
  \epsilon
  =
  \begin{pmatrix}
    \epsilon_{xx} & \epsilon_{xy} & 0 \\
    -\epsilon_{xy} & \epsilon_{yy} & 0 \\
    0 & 0 & \epsilon_{yy} \\
  \end{pmatrix}.
\end{equation}
This has an important implication. It means that
second order magnetic
contribution (which would appear as e.g.\ a difference between
$\epsilon_{yy}$ and $\epsilon_{zz}$) is either
absent in both systems or {\em not} resolvable with
standard electronic structure calculation. There is no reason why
second order magnetic contribution should be absent. So basically the
conclusion is that it is not resolvable with standard electronic structure
calculations. We discuss this in more detail in Sec.~\ref{sec:co}.

For the case of CrO$_2$ we conclude that
$\epsilon_{xz}$ is actually zero in the first order magnetic
contribution, however it might still be present in the third order.

\section{Polar MOKE of $\langle11\bar 20\rangle$ hcp Co} \label{sec:co}

\subsection{Optical conductivity}\label{sec:co_conductivity}

\begin{figure}
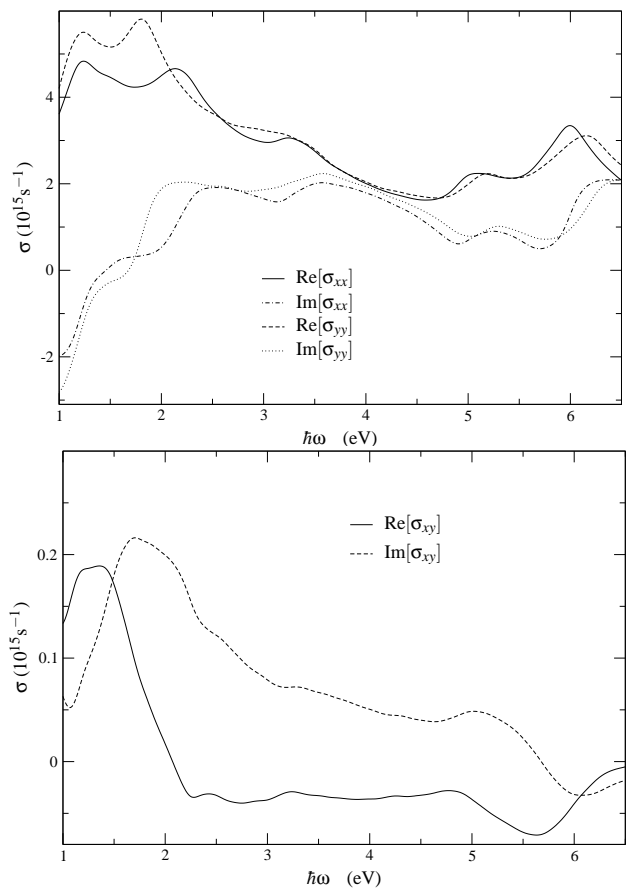

\begin{center}
  \psfrag{xtag}[c][b][.8]{$\hbar\omega$\hspace{2ex}(eV)}
  \psfrag{ytag}[c][t][.8]{$\sigma$ ($10^{15}$s$^{-1}$)}
  \psfrag{legend_re_xx}[l][l][.8]{Re$[\sigma_{xx}]$}
  \psfrag{legend_re_yy}[l][l][.8]{Re$[\sigma_{yy}]$}
  \psfrag{legend_im_xx}[l][l][.8]{Im$[\sigma_{xx}]$}
  \psfrag{legend_im_yy}[l][l][.8]{Im$[\sigma_{yy}]$}
  \includegraphics[scale=1,trim= 0cm 0cm 0cm 0cm,clip]{sig_diag_co.eps}\\
  \psfrag{xtag}[c][b][.8]{$\hbar\omega$\hspace{2ex}(eV)}
  \psfrag{ytag}[c][t][.8]{$\sigma$ ($10^{15}$s$^{-1}$)}
  \psfrag{legend_re_11-20}[l][l][.8]{Re$[\sigma_{xy}]$}
  \psfrag{legend_im_11-20}[l][l][.8]{Im$[\sigma_{xy}]$}
  \includegraphics[scale=1,trim= 0cm 0cm 0cm 0cm,clip]{sig_off_co.eps}\\
\caption{Calculated optical conductivity tensor of hcp Co with magnetisation direction
  $\langle 11\bar 20 \rangle$. Quantities are shown in a range where
  only direct inter band transition are important.}
\label{fig:sig_co}
\end{center}
\end{figure}

We have calculated the optical conductivity tensor of hcp
$\langle11\bar 20\rangle$ Co.
A hybridised
$4s4p3d$ and $5s5p4d$ basis was used in the calculations to
describe the Co atoms. Exchange correlation was taken into account
in the framework of the local spin density approximation in the
form proposed in Ref.~\onlinecite{vonbarth72jpc5:1629}.
The lattice constants were $a=2.5071\,${\AA} and
$c=4.0695\,${\AA}. 38400 k-points were used to sample the
Brillouin zone.
Results are shown in Fig.~\ref{fig:sig_co}. They are in good
agreement with previous theoretical results
\cite{guo94prb50:R10377,oppeneer95jmmm148:298}.
In the output of the calculation we find that tensor elements that
should be zero due to symmetry are of the order of $10^{9}\,$s$^{-1}$
while we find a
difference between $\sigma_{yy}$ and $\sigma_{zz}$ of the order of
$10^{13}..10^{12}\,$s$^{-1}$. Thus the symmetry of our calculated tensor
is in agreement with Eq.~(\ref{symmetry_full_co}) and
Eq.~(\ref{symmetry_expansion_co}). We conclude that the calculated difference
between $\sigma_{yy}$ and $\sigma_{zz}$ is a signature
of second order magnetic contribution. However if we change numerical
parameters of our calculation (e.\ g.\ basis set, k-point mesh) the
variation of $\sigma_{yy}$ and $\sigma_{zz}$ is typically larger. So
we have to conclude within the error of our calculation
$\sigma_{yy}$ and $\sigma_{zz}$ are equal. The conclusion is
that second order magnetic contribution can {\em not} be resolved with
standard electronic structure calculation.

\subsection{Optical response of the single crystal}

\begin{figure}
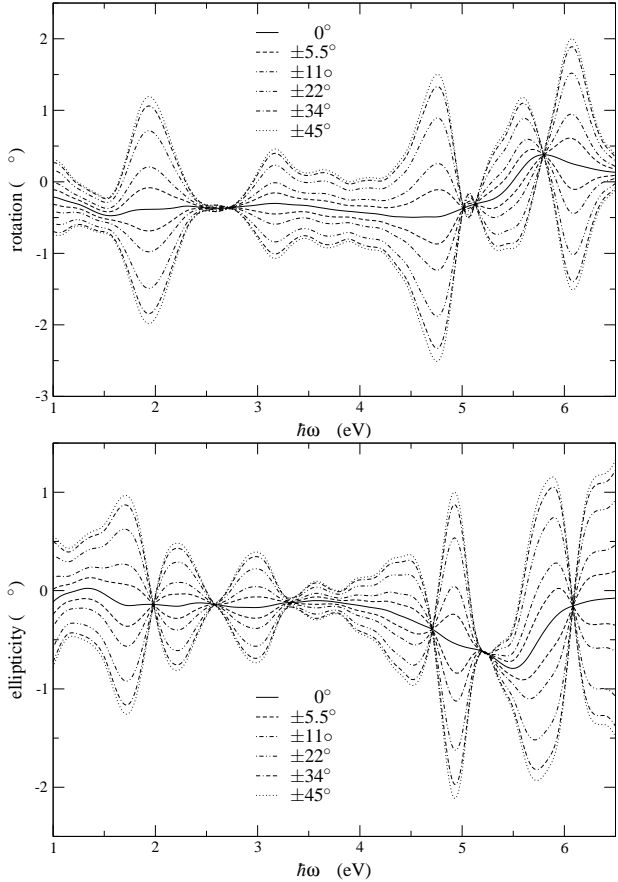

\begin{center}
  \psfrag{xtag}[c][b][.8]{$\hbar\omega$\hspace{2ex}(eV)}
  \psfrag{ytag}[c][t][.8]{rotation ($\phantom{12}^\circ$)}
  \psfrag{legend_0}[l][l][.8]{$\phantom{\pm 1}0^\circ$}
  \psfrag{legend_1o8}[l][l][.8]{$\pm 5.5^\circ$}
  \psfrag{legend_1o4}[l][l][.8]{$\pm 11\circ$}
  \psfrag{legend_1o2}[l][l][.8]{$\pm 22^\circ$}
  \psfrag{legend_3o4}[l][l][.8]{$\pm 34^\circ$}
  \psfrag{legend_1}[l][l][.8]{$\pm 45^\circ$}
  \includegraphics[scale=1,trim= 0cm 0cm 0cm 0cm,clip]{rot_full.eps}\\
  \psfrag{xtag}[c][b][.8]{$\hbar\omega$\hspace{2ex}(eV)}
  \psfrag{ytag}[c][t][.8]{ellipticity ($\phantom{12}^\circ$)}
  \psfrag{legend_0}[l][l][.8]{$\phantom{\pm 1}0^\circ$}
  \psfrag{legend_1o8}[l][l][.8]{$\pm 5.5^\circ$}
  \psfrag{legend_1o4}[l][l][.8]{$\pm 11\circ$}
  \psfrag{legend_1o2}[l][l][.8]{$\pm 22^\circ$}
  \psfrag{legend_3o4}[l][l][.8]{$\pm 34^\circ$}
  \psfrag{legend_1}[l][l][.8]{$\pm 45^\circ$}
  \includegraphics[scale=1,trim= 0cm 0cm 0cm 0cm,clip]{ellipt_full.eps}\\
\caption{Calculated optical response of $\langle 11\bar 20 \rangle$ hcp Co in
  polar MOKE geometry. The polarization vector is parallel to the
  crystallographic $x$-axis at zero angle. Curves shifted to higher
  values just below 5 eV correspond to positive angles.}
\label{fig:mono}
\end{center}
\end{figure}

We have calculated the optical
response in polar MOKE geometry with
perpendicular incident light with our transfer matrix approach. We
find that the optical response
depends strongly on the direction of the polarization vector in the
surface plane. If the
polarization vector is along one of the main crystal axis,
birefringence is absent and the
optical response is similar to common polar MOKE. When the polarization
vector is turned away from the main crystal axis the optical response is a
combination of crystallographic
birefringence and a magnetic effect. We find that birefringence starts to
be important at about $3^\circ$. Results are shown in 
Fig.~\ref{fig:mono}. Directions of the polarization vector are
in one quarter of the full circle in the surface plane which is
choosen symmetrically around the crystallographic $x$-axis. For directions of the
polarization vector chosen around the crystallographic $y$-axis,
results are identical on the scale of the plot. The latter is a non-trivial
result. Since the crystallographic $x$- and $y$-directions  are
different one would expect independent results in half of the full
circle. It can only be understood by stressing that the birefringence
is large compared to the magnetic effect (see below and
Sec.~\ref{sec:anisotropic}). The solid curves show
the case when the polarization vector is parallel to a main crystal axis. The
optical response is similar to the polar MOKE of hcp $\langle
0001 \rangle$ Co
\cite{guo94prb50:R10377,oppeneer95jmmm148:298}. To a good
approximation it can be regarded as a common polar MOKE response
without birefringence. The dashed and dotted curves show the
optical response for cases when
birefringence is important. If the polarization vector has an angle of
$\pm 5.5^\circ$ relative to the main crystal axis the birefringent
contribution has about the same magnitude as the magnetic
effect. It reaches its maximum at an angle of $\pm 45^\circ$. At this
angle it is about one order of magnitude larger than the
magnetic effect.

The present system has been investigated experimentally in detail by
Weller et al. \cite{weller94prl72:2097}. In this experiment different
samples were used at least one of which was a polycrystal with two
types of crystallographic domains related to
each other by a $90^\circ$ rotation around the 
surface normal. Experimental results do not report birefringent
contributions nor a dependence on the direction of the
polarization vector. So our theoretical results for the single crystal
presented here are very
different from experimental findings. Still, there is no direct
disagreement between theory and
experiment simply because it is possible that the experimental data
that was taken
actually corresponds to the case when the polarization vector is along a main crystal
axis. For this case there is good agreement with theory (see
Fig.~\ref{fig:comparison}). However we
believe that this is not what was happening. Rather we speculate that
during measurements at some point different directions of the
polarization vector were used and still basically common polar MOKE was found without
substantial dependence on the direction of the polarization
vector. Let us for the moment
focus on the sample which we know is a polycrystal. Then the
conclusion is optical response of a polycrystal with two domain
orientations is fundamentally different from the optical response of a
singlecrystal, so in order to describe experiment correctly it is
important to consider the full polycrystal rather than a
singlecrystalline sample.

\subsection{Optical response of the bicrystal}

We have calculated the optical response of a polycrystal with two
domain orientations. Our approach was to calculate average Stokes
parameters from our transfer matrix calculation as described in
Sec.~\ref{sec:ellipsometry}. Experimental data about the distribution of
domain sizes and intensities shining on them was not known so we had
to make an assumption here. We
expect that crystal growth occurrs with equal preferrence in both of the
two domain orientations so total surface areas should be the same and
total intensity of the incident light should be devided equally
among the two orientations.

\begin{figure}
  \psfrag{xtag}[c][b][.8]{$\hbar\omega$\hspace{2ex}(eV)}
  \psfrag{ytag}[c][t][.8]{polarization parameters ($\phantom{12}^\circ$)}
  \psfrag{psi}[c][t][.8]{$\psi$}
  \psfrag{chi}[c][t][.8]{$\chi$}
  \psfrag{legend_0}[l][l][.8]{$\phantom{+1}0^\circ$}
  \psfrag{legend_1o8}[l][l][.8]{$+22^\circ$}
  \psfrag{legend_3o8}[l][l][.8]{$-22\circ$}
  \psfrag{legend_1o16}[l][l][.8]{$+11^\circ$}
  \psfrag{legend_5o16}[l][l][.8]{$-11^\circ$}
  \includegraphics[scale=1]{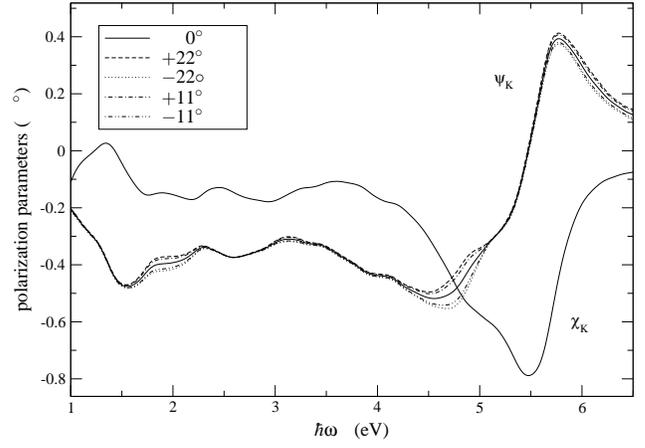}
  \caption{Calculated optical response of bicrystalline hcp
  $\langle 11\bar 20 \rangle$ Co. At 0$^\circ$ polarization vector is
  along a main crystal axis of one of the domains.}
  \label{fig:poly}
\end{figure}

Results are shown in Fig.~\ref{fig:poly}. In general, we find now for any direction of the
polarization vector that our calculated optical response is similar to common polar
MOKE and theoretical results are now in good
agreement with experimental data. The birefringent contribution, which
for the single crystal was the dominant contribution to the optical
response, is now averaged out. However birefringent contribution is
averaged out completely only in the ellipticity (in our computational result
variation under change of the direction of the polarization vector is
of the order $10^{-4\,\circ}$) while in the rotation it is still present.

In general the results are quite surprising: For the single crystal
birefringence was about 10 times larger than the
magnetic effect. For the polycrystal it is averaged out
so strongly that it is now smaller than
the magnetic contribution. How is this possible only due to the
presence of {\em one} additional domain
orientation? And secondly: why is the birefringent contribution
completely missing in the ellipticity but still present
in the rotation? It is important to find out the general mechanism
behind this.

We have considered average Stokes parameters for polycrystals with
ordered domains analytically. We find that the optical response
strongly depends on the in plane symmetry of the domain orientations.
In the majority of cases, ordered polycrystals
are equivalent to polycrystals with random domain distribution and thus
optical response is independent of the direction of the polarization
vector. In particular we can prove that the Stokes
parameters $S_0$ and $S_3$ are identical to those of a
random polycrystal if and only if the in plane symmetry of domain orientations is
larger than 2-fold and the Stokes parameters $S_1$ and $S_2$ are identical
to those of a random polycrystal if and only if the symmetry of domain
orientations is not 1, 2 or 4. The prove is given in the
appendix. Analytical findings are in good
agreement with the computational result we present here for the
hcp $\langle 11\bar 20 \rangle$ polycrystal with two domains. In particular
they explain the different behaviour of averaging
out in rotation and ellipticity (only $S_1$ and $S_2$
enter in the rotation, Eq.~\ref{rotation}, while mainly $S_3$ enters in the
ellipticity, Eq.~\ref{ellipticity}, now note the polycrystal with two domains
oriented by a $90^\circ$ rotation has 4-fold symmetry). The
analytical findings have an important consequence for
experiments. They imply that if only a few ordered domains are present inside
the illuminated area the optical response will always be very close to
common polar MOKE.

\begin{figure}
\begin{center}
  \psfrag{xtag}[c][b][.8]{$\hbar\omega$\hspace{2ex}(eV)}
  \psfrag{ytag}[c][t][.8]{rotation ($\phantom{12}^\circ$)}
  \psfrag{legend_poly}[l][l][.8]{$\langle11\bar2 0\rangle$ polycrystal
  $p||x$}
  \psfrag{legend_x}[l][l][.8]{$\langle11\bar2 0\rangle$ single crystal $p||x$}
  \psfrag{legend_y}[l][l][.8]{$\langle11\bar2 0\rangle$ single crystal $p||y$}
  \psfrag{legend_0001}[l][l][.8]{$\langle0001\rangle$}
  \psfrag{legend_weller_11-20}[l][l][.8]{$\langle11\bar 20\rangle$ experiment}
  \psfrag{legend_weller_0001}[l][l][.8]{$\langle0001\rangle$ experiment}
  \includegraphics[scale=1,trim= 0cm 0cm 0cm 0cm,clip]{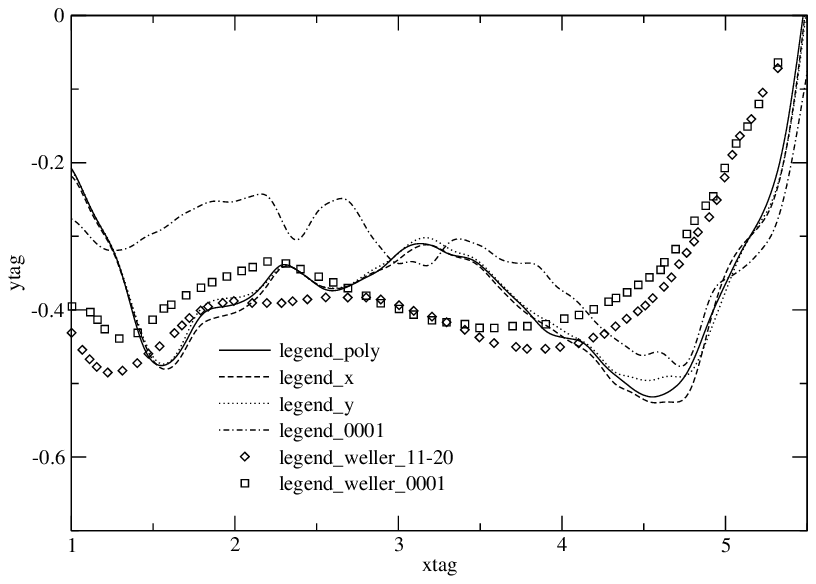}\\
  \psfrag{xtag}[c][b][.8]{$\hbar\omega$\hspace{2ex}(eV)}
  \psfrag{ytag}[c][t][.8]{ellipticity ($\phantom{12}^\circ$)}
  \psfrag{legend_poly}[l][l][.8]{$\langle11\bar2 0\rangle$ polycrystal
  $p||x$}
  \psfrag{legend_x}[l][l][.8]{$\langle11\bar2 0\rangle$ single crystal $p||x$}
  \psfrag{legend_y}[l][l][.8]{$\langle11\bar2 0\rangle$ single crystal $p||y$}
  \psfrag{legend_0001}[l][l][.8]{$\langle0001\rangle$}
  \psfrag{legend_weller_11-20}[l][l][.8]{$\langle11\bar 20\rangle$ experiment}
  \psfrag{legend_weller_0001}[l][l][.8]{$\langle0001\rangle$ experiment}
  \includegraphics[scale=1,trim= 0cm 0cm 0cm 0cm,clip]{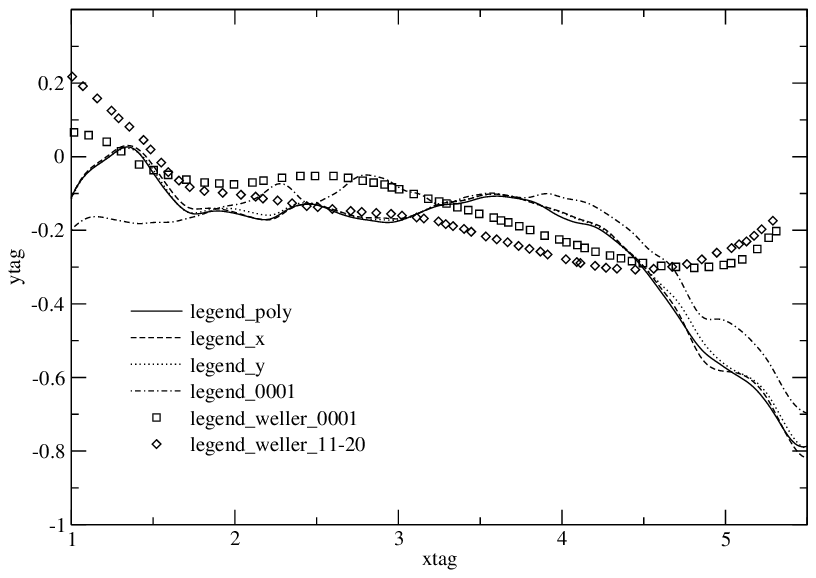}\\
\caption{Optical response of hcp $\langle 11\bar 20 \rangle$ Co in
  polar MOKE geometry. Theoretical data is for special cases of the direction of the
  polarization vector. Experimental data is due to Weller et al. Results for
  hcp $\langle0001\rangle$ are shown for comparison.}
\label{fig:comparison}
\end{center}
\end{figure}

We now conjecture that the second sample that was
investigated in experiments (the Ru($11\bar 20$) sample) was
also a polycrystal (the presence of few ordered domains in the
illuminated area is enough). For any direction of the
polarization vector that was possibly considered in experiment we
immediately have agreement with theory. Summarizing comparison of
theoretical data with experiment is shown in
Fig.~\ref{fig:comparison}. Data for $\langle 0001\rangle$ hcp Co are
shown for comparison. The theoretical data for $\langle
0001\rangle$ hcp Co has been calculated in the same way as the data
for $\langle 11\bar 20 \rangle$ hcp. It is in good agreement with
previous theoretical data \cite{oppeneer95jmmm148:298,guo94prb50:R10377}.

\subsection{Anisotropic polar MOKE}\label{sec:anisotropic}

The goal of the previous experimental work of Weller and coworkers was
to find a manifestation of magneto-crystalline anisotropy in the
magneto-optical response. They investigated how the optical response
changes when the relative orientation between magnetization
and crystal lattice is changed while the polar measuring geometry
as well as other parameters of the experiment are kept
(lattice parameters, crystal growth quality, etc.). 
It was found that the
optical response of hcp $\langle 0001 \rangle$ and hcp $\langle
11\bar 2 0\rangle$ is different. These results were
explained by the dependence of the absorptive part of the refractive
index on the angle between crystallographic $c$-axis and spin moment.

We know now that the electrodynamics part of the problem is much more
complicated. It is important to calculate the full optical response
including crystallographic birefringence and also the
polycrystalline nature of the sample has to be taken into account. So it is
important to check if the main conlucions
given in the experimental work stil hold. As we will see below, the answer is
yes.

From a theoretical point of view the situation is the following: We have
common polar MOKE in hcp $\langle0001\rangle$ (no birefringence,
optical response is independ of direction of polarization vector) and
a combination of birefringence and magnetic response with strong
averaging out of birefringence in the hcp $\langle11\bar 20\rangle$
polycrystal. So optical responses are fundamentally
different. Nevertheless in both systems the magnetic contribution to the optical
response originates from the tensor element $\epsilon_{xy}$. We would say
that we have measured anisotropy in the magneto-optical constants
if we can conclude from the measurement that $\epsilon_{xy}$ has
changed due to change of the magnetization direction. So what we want
to show now is that the difference in
magneto-optical response between single crystalline hcp $\langle0001\rangle$ and
polycrystalline hcp $\langle11\bar 20\rangle$ is basically only determined by
the change in $\epsilon_{xy}$. Admittedly we do not think this can be proven
rigorously, however what we can do is to calculate the optical
response of the hcp $\langle11\bar 20\rangle$ crystal with a
dielectric tensor were we substitute $\epsilon_{yy}$ by $\epsilon_{xx}$
and vice versa. We can also use the average
$\frac{1}{2}(\epsilon_{xx}+\epsilon_{yy})$ for both. We find in any
case the optical response is very close to both the result obtained for the
single crystal with the polarization vector along a main crystal axis and
for the polycrystal. All these cases are much closer to each other
than to the result for hcp $\langle0001\rangle$; see also
Fig.~\ref{fig:comparison}. The conclusion is that the difference between
optical response of hcp $\langle11\bar 20\rangle$ and hcp
$\langle0001\rangle$ is mainly due to the change in
$\epsilon_{xy}$. In this sense it may be regarded as anisotropic polar
MOKE or a manifestation of magneto-crystalline anisotropy in the
optical response.

\section{Polar MOKE of {CrO$_2$}}\label{sec:cro2}

\subsection{Optical conductivity}\label{sec:cro2_conductivity}

We have calculated the optical conductivity tensor of $\langle 010
\rangle$ CrO$_2$ with the first principles approach as described in
Sec.~\ref{sec:first_principles}. The basis set was constructed from
$4s4p3d$ and $4d4f$, (respectively $2s2p$ and $3s3p$) orbitals
for the chromium (respectively oxygen) sites. The lattice
constants and position parameters were {$a=4.421\,$\AA}, {$c=2.916\,$\AA} and
$x=0.3053$ as it was used in Refs. \onlinecite{thamer57jacs79_547,
schwarz86jpf16_211,matar92jpi2_315,uspenskii96prb54:474}.
32768 k-points were used to sample the Brillouin zone. Exchange
correlation was treated in the same way as in the calculation for Co
above.

\begin{figure}
  \psfrag{xtag}[c][b][.8]{$\hbar\omega$\hspace{2ex}(eV)}
  \psfrag{ytag}[c][t][.8]{$\sigma_{xx}$\hspace{2ex}($10^{15}\,$s$^{-1}$)}
  \psfrag{legend_re_xx}[l][l][.8]{Re$[\sigma_{xx}]$}
  \psfrag{legend_re_yy}[l][l][.8]{Re$[\sigma_{yy}]$}
  \psfrag{legend_im_xx}[l][l][.8]{Im$[\sigma_{xx}]$}
  \psfrag{legend_im_yy}[l][l][.8]{Im$[\sigma_{yy}]$}
  \includegraphics[scale=1,trim= 0cm 0cm 0cm 0cm,clip]{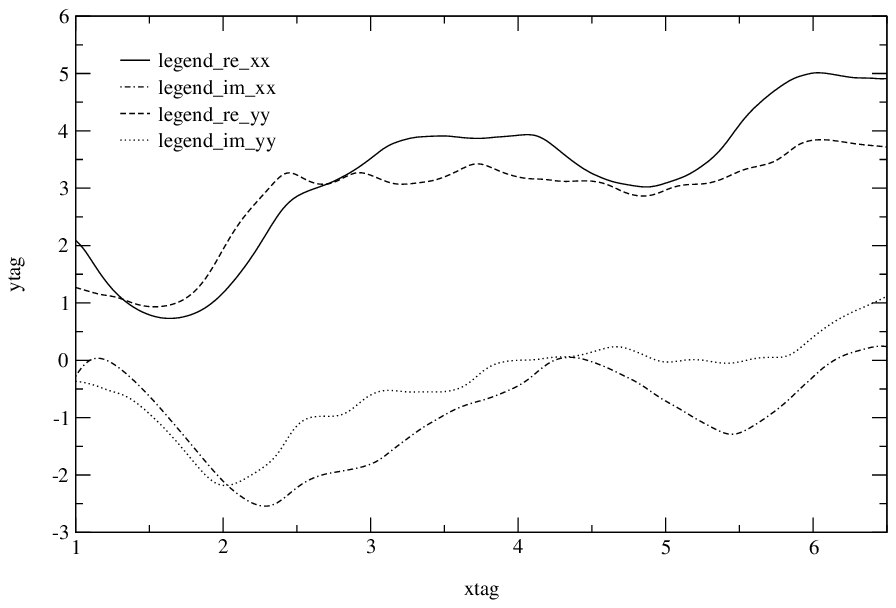}\\
  \psfrag{ytag}[c][t][.8]{$\sigma_{xy}$\hspace{2ex}($10^{15}\,$s$^{-1}$)}
  \psfrag{legend_re}[l][l][.8]{Re$[\sigma_{xy}]$}
  \psfrag{legend_re}[l][l][.8]{Re$[\sigma_{xy}]$}
  \includegraphics[scale=1,trim= 0cm 0cm 0cm 0cm,clip]{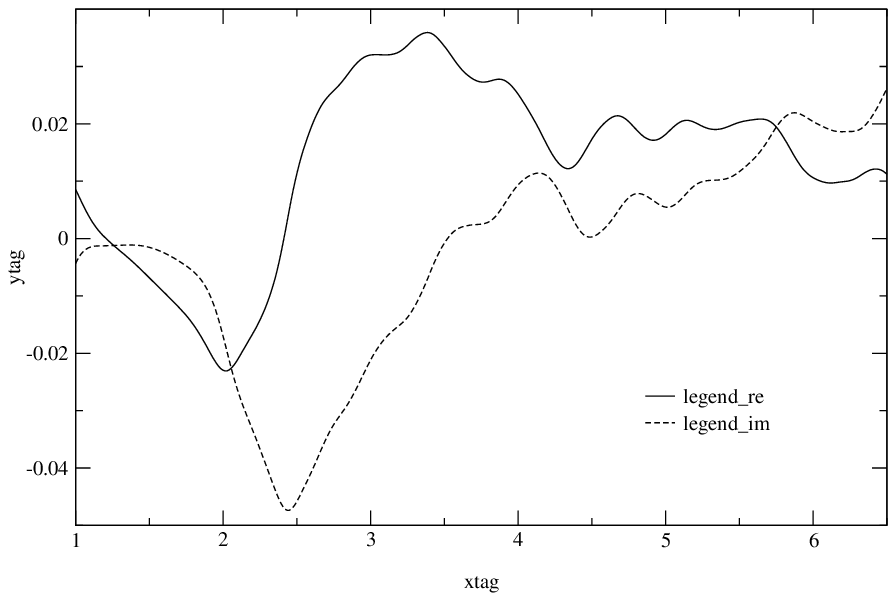}\\
  \caption{Calculated optical conductivity tensor of $\langle 010 \rangle$ CrO$_2$.}
  \label{fig:sig_cro2}
\end{figure}

The magnetic moment per CrO$_2$ $m=2.0\,\mu_B$ and total energy
per unit cell as well as the DOS agree well with those given in
Refs.~\onlinecite{matar92jpi2_315,uspenskii96prb54:474,kunes02prb65:165105}.
Fig.~\ref{fig:sig_cro2} shows our calculated optical conductivity tensor. Results
are in good agreement with previous theoretical findings
\cite{uspenskii96prb54:474,kunes02prb65:165105}.

\subsection{Optical response of the polycrystal}

If thin films of CrO$_2$ are deposited on single-crystalline
Al$_2$O$_3$, polycrystalline growth is observed. Crystallites
order \mbox{6-fold} symmetrically with an $a$-axis oriented
perpendicular to the surface \cite{rabe02jpcm14:7}. Experimental
results suggest that the sizes of crystallites in
such films are typically of the order $0.1$--$10\,\mu$m. For the lower
limit we are in a regime where interference effects start to
play a role. Consequently the optical response is no longer a
purely incoherent wave and can in general {\em
  not} be described by average Stokes
parameters. We exclude this case
here. For the upper limit the optical response of the
polycrystal is well described by average Stokes parameters.

\begin{figure}
  \psfrag{xtag}[c][b][.8]{$\hbar\omega$\hspace{2ex}(eV)}
  \psfrag{ytag}[c][t][.8]{Kerr parameters ($\phantom{12}^\circ$)}
  \psfrag{psi}[c][t][.8]{$\psi$}
  \psfrag{chi}[c][t][.8]{$\chi$}
  \includegraphics[scale=1,trim= 0cm 0cm 0cm 0cm,clip]{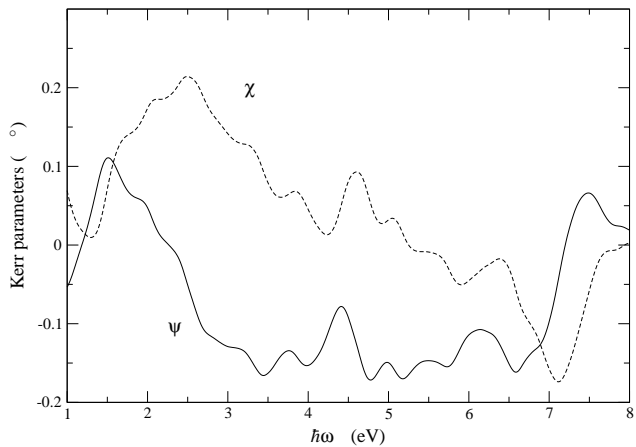}\\
  \caption{Calculted optical response of $a$-axis textured 6-fold
  symmetrically ordered CrO$_2$. The solid line shows the
  rotation $\psi$ of the polarization ellipse. The dashed line shows
  its ellipticity $\chi$.}
  \label{fig:cro2_moke}
\end{figure}

We have calculated the optical response of polycrystalline $\langle
010 \rangle$ CrO$_2$ with $6$-fold symmetric domain ordering. We find
that the optical response is independent of the direction of the
polarization vector. Results are shown in
Fig.~\ref{fig:cro2_moke}. They are in good agreement with
experimental data.

Also, results are in good agreement with analytical findings given in
appendix \ref{app:classification}: the $6$-fold symmetric polycrystal
is a member of the isotropic class which implies that crystallographic
birefringence is averaged out completely both in the rotation and in
the ellipticity.

Results have an important implication. In a previous theoretical
work Uspenski{\v i} et al.\cite{uspenskii96prb54:474} derived an
approximative analytic expression for the polar MOKE of a polycrystalline
surface with two-dimensional random domain distribution. It reads
\begin{equation}\label{uspenskii}
  \psi+i\chi
  =
  \frac{2\epsilon_{xy}}{(\sqrt{\epsilon_{xx}}+
  \sqrt{\epsilon_{yy}})(1-\sqrt{\epsilon_{xx}}\sqrt{\epsilon_{yy}})}.
\end{equation}
Here the roots are taken in the upper complex half plane. From
the more recent experimental works \cite{rabe02jpcm14:7} it is clear
that domain distribution of
polycrystalline CrO$_2$ is actually {\em not} random rather it has
$6$-fold symmetry. So Eq.~(\ref{uspenskii}) is in general not
applicable. However, from the analytical results of
appendix~\ref{app:classification} we know now that the optical response of
the $6$-fold symmetric polycrystal is equivalent to the optical
response of a random polycrystal of the same material. Thus,
the validity of the approximative expression is extended to the whole
isotropic class. Hence, indeed the optical response of CrO$_2$ can be
calculated with Eq.~(\ref{uspenskii}).

We have calculated the optical response also with the
approximative expression. Results differ from the rigorous result
obtained with our transfer matrix calculation and subsequent determination
of exact average Stokes parameters in the fourth relevant digit. This
shows that (for CrO$_2$) the approximative expression is actually very
good. Also it shows that computational results are in very good
agreement with the rigorous analytic treatment given in the appendix.

\section{Summary and conclusion}\label{sec:conclusions}

We have calculated the polar magneto-optical Kerr effect for hcp
$\langle11\bar 20\rangle$ Co and for $\langle 010 \rangle$
CrO$_2$. Our approach was based on first-principle calculations of dielectric tensors.
We have addressed the electrodynamics part of the problem,
i.~e., the extraction of MOKE from dielectric tensors, with a transfer
matrix method. We could describe simultaneous occurrence of birefringence
and magnetic effect that is present in the systems. For
polycrystals average optical response was described by exact average
Stokes parameters taking into account the real orientations of domains.

For hcp $\langle11\bar 20\rangle$ Co we found
that a single crystal optical response depends strongly on
the direction of the polarization vector. If
the polarization vector is along one of the main crystal axis optical
response is very similar to common polar MOKE and moreover for the two
crystal axis the optical response is basically the same. If the
polarization vector deviates more than about $3^\circ$ from one of the
main crystal axis birefringence is important. For larger angles it
dominates over the actual magnetic effect. To explain
experimental data we had to stress that samples investigated
in experiment were polycrystals. We could show that already the
presence of two domain orientations leads to a strong reduction of
birefringent contribution in the magneto-optical response. Finally we
could show that the previous interpretation of experimental data in
terms of a manifestation of magneto-crystalline anisotropy in the
optical response remains valid.

For polycrystalline $\langle 010 \rangle$ CrO$_2$ we found
that the birefringent contribution to the optical response is averaged out
completely. We could verify that a previous approximative analytic
expression describes the optical response exactly also for the
case of realistic domain orientations.

The results of our LDA calculations for both hcp Co and CrO2 are in very good agreement
with the experimental data (assuming that the data for Co are for a
bi-crystal). This is not trivial since, in
general, correlation effects might be essential for
the electronic structure of transition metal ferromagnets
\cite{lichtenstein01prl87:067205}. The effect of local Coulomb
interactions
on magneto-optical properties of Fe and Ni has been calculated in
Refs.~\onlinecite{perlov04jmmm272-276:523,perlov04physics_of_spin_in_solids:161}
in a framework
of dynamical mean-field theory (LDA+DMFT approach). It appeared that, whereas for Ni
the correlation effects are important, for Fe there are almost no difference between LDA
and LDA+DMFT results for optical and magneto-optical properties. Our results show
that probably correlation effects are not very important also for magneto-optical properties
of Co. As for ferromagnetic CrO$_2$ recent analysis
\cite{toropova04cond-mat:0409554} shows that it should be considered
rather as a weakly correlated system so a success of our calculations
is not surprising.

\begin{acknowledgments}
We gratefully acknowledge stimulating discussions with Thorsten
Andersen, Peter Oppeneer and Till Burkert. This work was supported financially by the
Deutscher Akade\-mi\-scher Austauschdienst, Germany, by the Studienstiftung der
Deutschen Wirtschaft, Studienf\"orderwerk Klaus Murmann, Germany, by the
Center for Dynamical Processes, University of Uppsala, Sweden, by the
Swedish Research Council (VR), the foundation for strategic research
(SSF) and the G\"oran Gustafsson foundation.
\end{acknowledgments}

\appendix

\section{Classification of polycrystalline surfaces}\label{app:classification}

Most polycrystalline surfaces occurring in nature have either a
three-dimensional distribution of domain orientations or a
two-dimensional distribution with only few domain orientations
that are related to
each other by a rotation round the surface normal.
Three-dimensional distribution is found for surfaces of
bulk polycrystals such as, e.g., natural iron. Ordered
two-dimensional distribution is often found when thin
polycrystalline films are grown on single-crystalline substrates.
In the case
of three-dimensional distribution, the domain orientations are
often to a good approximation random. The average polar MOKE of a
three-dimensional random polycrystal is obviously independent of the
direction of the polarization vector
in the surface plane. We skip this case here as well as other
three-dimensionally ordered polycrystals. Rather we focus on
polycrystals with a two-dimensional distribution of domain
orientations.  We call
a surface $n$-fold symmetrically
ordered if the crystallographic structures of all domains can be
mapped onto each other by an $n$-fold rotation around the surface
normal. We will also use a notion of two-dimensional continuously
distributed polycrystalline surface. By that we mean a
polycrystalline
surface in which the crystallographic structures of the domains
can be mapped onto each other by suitable continuous rotations
around the surface normal and all possible orientations occur.
This corresponds to two-dimensional random domain orientations.
Also for this case, the average polar MOKE is obviously independent of
the direction of the polarization vector.

We show now that for most polycrystals with
symmetrically ordered domains the average polar MOKE is equivalent to
the average polar MOKE of a continuously distributed polycrystal of
the same material.

In particular we prove the following statement. The average Stokes
parameters $\langle S_0\rangle$ and $\langle S_3\rangle$ are identical
to those of a continuous polycrystal if and only if the in plane
symmetry of domain orientations is larger than two-fold and the
Stokes parameters $\langle S_1\rangle$ and $\langle S_2\rangle$ are identical
to those of a continuous polycrystal if and only if the in plane
symmetry of domain orientations is not $1$, $2$ or $4$.

We begin the proof by considering the light reflected from a
single domain. If reflection is described by means of transfer matrix
method, then, for any wave
vector and frequency, the complex amplitude of the reflected wave
is a linear mapping of the complex amplitude of the incident wave.
This can be seen directly from the main linear equation, Eq.~(\ref{linear}) .
Further, in case of normal incidence, the incident and the
reflected amplitude vectors may be represented in a common
coordinate system parallel to the surface plane. Thus, if $\vec
E^{in}$ and $\vec E^{refl}$ are respective \mbox{$2$-vectors},
there is a linear transformation
\mbox{$T:\mathbb{C}^2\rightarrow\mathbb{C}^2$} such that
\begin{equation}
  \vec E^{refl}=T\vec E^{in}.
\end{equation}
Now let some other domain be identical to the previous one up to a
rotation
\begin{equation}
  R(\varphi)
  =
  \begin{pmatrix}
    \cos\varphi & \sin\varphi \\
    -\sin\varphi & \cos\varphi \\
  \end{pmatrix}
\end{equation}
around the surface normal.

If $\vec E_2^{refl}$ is the amplitude of the wave reflected from
the second domain, we have
\begin{equation}
  \vec E_2^{refl}=R(\varphi)TR^{-1}(\varphi)\vec E^{in}.
\end{equation}
Now consider an $n$-fold symmetrically ordered polycrystal.
Introducing angles \mbox{$\varphi_k=2\pi\frac{k}{n}$},
\mbox{$k=1,\dots,n$}, the amplitude vectors $\vec E_k^{refl}$ of
the reflected waves are
\begin{equation}
  \vec E_k^{refl}=R(\varphi_k)TR^{-1}(\varphi_k)\vec
  E^{in}.
\end{equation}
By Eq.~(\ref{stokes_parameters}) the average Stokes parameters are
\begin{equation}
  \langle\vec S\rangle=\sum_{k=1}^n\vec
  S\left(R(2\pi\frac{k}{n})TR^{-1}(2\pi\frac{k}{n})\vec E^{in}
  \right).
\end{equation}
For a continuously distributed polycrystal we have
\begin{equation}
  \langle\vec S\rangle=\frac{1}{2\pi}\int_{0}^{2\pi} \vec S
  \left(R(\varphi)TR^{-1}(\varphi) \vec E^{in}\right)\,d\varphi.
\end{equation}
It is shown in appendix \ref{app:polycrystal} that the latter two
expressions are equal in the first and last component if and only
if \mbox{$n\not\in\{1,2\}$} and in the second and third component
if and only if \mbox{$n\not\in\{1,2,4\}$}. This finishes the
proof.

The latter statement is fundamental for the understanding of the
average polar MOKE of polycrystals. It naturally
decides thin polycrystalline films into three classes:
\mbox{2-fold} symmetrically ordered films, \mbox{4-fold}
symmetrically ordered films and all others including
two-dimensional random orientation. Further, it implies that
for polycrystalline films out of the first two classes the
optical response does in general
depend on the direction of the polarization vector, thus the
birefringent contribution to the optical response is in general {\em not}
averaged out. On the other hand it implies that for
polycrystals out of the last class optical response is independent
of the direction of the polarization vector, thus the birefringent
contribution to the optical response {\em is} averaged out. 

\section{Symmetric sums over powers of trigonometric functions}

We prove a statement about symmetric sums over powers of $\cos$
and $\sin$.

Let \mbox{$f:\mathbb{R}\rightarrow\mathbb{R}$} and
\mbox{$q\in\mathbb{Q}$}. Then the identity
\begin{equation*}
  \frac{1}{2\pi}\int_0^{2\pi} f(x)\,dx=
    \frac{1}{n}\sum_{k=1}^n f({\textstyle2\pi\frac{k}{n}})=q
\end{equation*}
holds for pairs $f$, $q$
\begin{equation}\label{n_2}
\begin{aligned}
    &\cos^2\,\,,\quad\frac{1}{2}\\
    &\sin^2\,\,,\quad\frac{1}{2}\\
    &\cos\,\sin\,\,,\quad 0\\
    &\cos^4+\cos^2\,\sin^2\,\,,\quad\frac{1}{2}\cdot\frac{3}{4}+\frac{1}{8}\\
    &\sin^4+\cos^2\,\sin^2\,\,,\quad\frac{1}{2}\cdot\frac{3}{4}+\frac{1}{8}\\
    &\cos^3\sin\,+\sin^3\cos\,\,,\quad0\\
    &\sin^4\,-\cos^4\,\,,\quad0\\
\end{aligned}
\end{equation}
if and only if $n\not\in\{1,2\}$, and for pairs $f$, $q$
\begin{equation}\label{n_4}
\begin{aligned}
    &\cos^4\,\,,\quad\frac{1}{2}\cdot\frac{3}{4}\\
    &\sin^4\,\,,\quad\frac{1}{2}\cdot\frac{3}{4}\\
    &\cos^2\,\sin^2\,\,,\quad\frac{1}{8}\\
    &\cos^3\,\sin\,\,\,,\quad0\\
    &\sin^3\,\cos\,\,\,,\quad0\\
    &\cos^4-\cos^2\,\sin^2\,\,,\quad\frac{1}{2}\cdot\frac{3}{4}-\frac{1}{8}\\
    &\sin^4-\cos^2\,\sin^2\,\,,\quad\frac{1}{2}\cdot\frac{3}{4}-\frac{1}{8}\\
    &\cos^3\sin\,-\sin^3\cos\,\,,\quad0\\
\end{aligned}
\end{equation}
if and only if $n\not\in\{1,2,4\}$.

We begin the proof by considering sums of the form
\begin{equation}\label{general_sum}
  \frac{1}{n}\sum_{k=1}^n e^{mi2\pi \frac{k}{n}}\,\,,\quad
  m\in\mathbb{N},
\end{equation}
where $n\in \mathbb{N}$, $n\ge 2$.

If $m=l\,n$ with some $l\in\mathbb{N}$, then
\begin{equation}\label{sum_smaller}
  \frac{1}{n}\sum_{k=1}^n e^{mi2\pi\frac{k}{n}} = \frac{1}{n}\sum_{k=1}^n
  e^{li2\pi k} = 1,
\end{equation}

whereas if $n=l\,m$ with some $l\in\mathbb{N}$ we have
\begin{equation}\label{sum_greater}
  \frac{1}{n}\sum_{k=1}^{n} e^{mi2\pi\frac{k}{n}} = \frac{1}{n}\sum_{k=1}^{lm}
  e^{i2\pi\frac{k}{l}} = \frac{1}{n}\,m\sum_{k=1}^{l} e^{i2\pi\frac{k}{l}} = 0.
\end{equation}

Now let $p$ and $q$ be the largest prime numbers occurring in the
prime factorisations of $n$ and $m$ respectively. Let
\mbox{$\mathbb{F}_p=\left\{\left\{0\,,\,1\,,\dots,\,p-1\right\},\cdot\,,+\right\}$}
be the prime field of the modulo classes of $p$ in the common
sense. Let $m\cdot\mathbb{F}\subset\mathbb{N}$ be the set
\mbox{$\{0\,,\,1q\diagdown
  p\,,\,2q\diagdown p\,,\,\dots\,,\,(p-1)q\diagdown p\}$}. We devide the set
of complex numbers occurring in Eq.~(\ref{general_sum}) into
\mbox{$s$ subsets}. We \mbox{chose $s$} such that \mbox{$n=s\cdot
p$} and consider
\begin{equation}
  A_j=\left\{e^{i2\pi\frac{k}{p}+mi2\pi\frac{j}{n}}, k\in
  m\cdot\mathbb{F}_{p} \right\}\,\,,\quad j=1,\dots,s.
\end{equation}
Then
\begin{equation}\label{sum_prime}
\begin{split}
  \frac{1}{n}\sum_{k=1}^n e^{mi2\pi \frac{k}{n}}
  =
  \frac{1}{n}\left[
    \sum_{z\in A_1}
    +\dots+
    \sum_{z\in A_s}
  \right]\\
  =
  \frac{1}{n}\left[
    \left(  e^{mi2\pi\frac{1}{n}}+\dots+e^{mi2\pi\frac{s}{n}}  \right)
    \sum_{k\in m\cdot\mathbb{F}_p}e^{i2\pi\frac{k}{p}}
  \right]
  = 0 \quad\text{if}\quad q<p.
\end{split}
\end{equation}

Using Eq.~(\ref{sum_smaller}), Eq.~(\ref{sum_greater}) and
Eq.~(\ref{sum_prime}), we can calculate the sum given by
Eq.~(\ref{general_sum}) with some \mbox{$m\in\mathbb{N}$} for any
\mbox{$n\in\mathbb{N}$}. We consider the cases $m=2$ and $m=4$.

For $m=2$ we obviously have a largest prime factor $q=2$, i.e., by
Eq.~(\ref{sum_prime}) the sum vanishes for
\begin{equation*}
  n=3,6,7,9,10,11,12,\dots
\end{equation*}
and any other natural number containing a prime greater than or
equal to three in its factorisation. \mbox{For $n=4,8,16,\dots$},
the sum vanishes by Eq.~(\ref{sum_greater}), while for
\mbox{$n=1,2$} the sum is one by Eq.~(\ref{sum_smaller}). Thus we
have
\begin{equation}\label{sum_n_2}
  \frac{1}{n}\sum_{k=1}^n e^{2i2\pi \frac{k}{n}}
  = 0\quad\text{if and only if}\quad n\not\in\{1,2\}.
\end{equation}

For $m=4$, we have a largest prime factor $q=2$ as well, i.e., the
sum vanishes again for
\begin{equation*}
  n=3,6,7,9,10,11,12,\dots
\end{equation*}
and any other natural number containing a prime greater than or
equal to three in its factorisation. \mbox{For $n=8,16,32,\dots$},
the sum vanishes by Eq.~(\ref{sum_greater}), while for
\mbox{$n=1,2,4$}, we obtain by Eq.~(\ref{sum_smaller}) that the
sum is one. Thus
\begin{equation}\label{sum_n_4}
  \frac{1}{n}\sum_{k=1}^n e^{4i2\pi \frac{k}{n}}
  = 0\quad\text{if and only if}\quad n\not\in\{1,2,4\}.
\end{equation}

We prove the first line of Eq.~(\ref{n_2}). The identity
\begin{equation}
  \frac{1}{2\pi}\int_0^{2\pi}\cos^2(\varphi)\,d\varphi=\frac{1}{2}
\end{equation}
follows from the more general formula
\begin{equation}\label{cos_sin_int}
  \int_0^{\pi/2}\sin^{2\alpha+1}(\varphi)\cos^{2\beta+1}(\varphi)\,d\varphi
  =\frac{\Gamma(\alpha+1)\Gamma(\beta+1)}{2\Gamma(\alpha+1+\beta+1)},
\end{equation}
where $\Gamma$ is the gamma function
\cite{bronstein99taschenbuch_der_mathematik}. Further,
\begin{equation}
\begin{aligned}
  \frac{1}{n}\sum_{k=1}^n\cos^2({\textstyle2\pi\frac{k}{n}})
  &=
  \frac{1}{2} + \frac{1}{n}\cdot\frac{1}{4}\sum_{k=1}^n\left[ e^{2i2\pi\frac{k}{n}}
  + e^{-2i2\pi\frac{k}{n}} \right]\\
  &=\frac{1}{2} \quad\text{if and only if}\quad n\not\in\{1,2\},
\end{aligned}
\end{equation}
where we have used Eq.~(\ref{sum_n_2}) for the last line.

In a similar way, the second and third line of Eq.~(\ref{n_2})
follow from Eq.~(\ref{cos_sin_int}) and Eq.~(\ref{sum_n_2}).

Next we prove the first line of Eq.~(\ref{n_4}). Once again, we
refer to Eq.~(\ref{cos_sin_int}) to see that
\begin{equation}
  \frac{1}{2\pi}\int_0^{2\pi}\cos^4(\varphi)\,d\varphi=\frac{1}{2}\cdot\frac{3}{4}.
\end{equation}
On the other hand
\begin{equation}
\begin{aligned}
  \frac{1}{n}&\sum_{k=1}^n\cos^4({\textstyle2\pi\frac{k}{n}})\\
  &=
  \frac{1}{2}\cdot\frac{3}{4} + \frac{1}{n}\cdot\frac{1}{16}
  \sum_{k=1}^n\left[ e^{4i2\pi\frac{k}{n}} +
  e^{-4i2\pi\frac{k}{n}} + 4e^{2i2\pi\frac{k}{n}} +
  4e^{-2i2\pi\frac{k}{n}}\right]\\
  &=
  \frac{1}{2}\cdot\frac{3}{4} \quad\text{if and only if}\quad n\not\in\{1,2,4\},
\end{aligned}
\end{equation}
where we have used Eq.~(\ref{sum_n_2}) and Eq.~(\ref{sum_n_4}) for
the last line.

In a similar way, we find
\begin{equation}
  \frac{1}{n}\sum_{k=1}^n\sin^4({\textstyle2\pi\frac{k}{n}})
  =
  \frac{1}{2}\cdot\frac{3}{4},
\end{equation}
\begin{equation}
  \frac{1}{n}\sum_{k=1}^n\cos^2({\textstyle2\pi\frac{k}{n}})
     \sin^2({\textstyle2\pi\frac{k}{n}})
  =
  \frac{1}{8},
\end{equation}
\begin{equation}
  \frac{1}{n}\sum_{k=1}^n\cos^3({\textstyle2\pi\frac{k}{n}})
    \sin({\textstyle2\pi\frac{k}{n}})
  =
  0
\end{equation}
and
\begin{equation}
  \frac{1}{n}\sum_{k=1}^n\cos({\textstyle2\pi\frac{k}{n}})
    \sin^3({\textstyle2\pi\frac{k}{n}})
  =
  0
\end{equation}
if and only if $n\not\in\{1,2,4\}$.
This gives the first five identities of Eq.~(\ref{n_4}).

To see that the last four lines of Eq.~(\ref{n_2}) hold, we add
the corresponding expressions obtained above and find that in all
cases the sums over fourth powers cancel, while sums over second
powers remain. In contrast to that, also the fourth order sums
remain in the expressions for the last three lines of
Eq.~(\ref{n_4}). This finishes the proof.

\section{Average Stokes parameters for $n$-fold
rotated $2\times 2$ linear transformations}\label{app:polycrystal}

Let $\vec E\in\mathbb{R}^2$,
$R(\varphi):\mathbb{R}^2\rightarrow\mathbb{R}^2$ be a rotation by
an angle $\varphi$ and $T:\mathbb{C}^2\rightarrow\mathbb{C}^2$ a
linear transformation of most general symmetry. Let
$S_j:\mathbb{C}^2\rightarrow\mathbb{R}$, $j=0,1,2,3$ be the Stokes
parameters. Then
\begin{equation}\label{stokes_equal}
\begin{aligned}
  \frac{1}{n}\sum_{k=1}^n
    &S_j\left(R({\textstyle2\pi\frac{k}{n}})T
    R({\textstyle2\pi\frac{k}{n}})^{-1}\vec E\right)\\
  &=\frac{1}{2\pi}\int_0^{2\pi}S_j\left(R(\varphi)TR^{-1}(\varphi)\vec
    E\right)\,d\varphi
\end{aligned}
\end{equation}
holds for $j=0,3$ if and only if $n\not\in\{1,2\}$ and for $j=1,2$
if and only if $n\not\in\{1,2,4\}$.\vspace{.3cm}

We treat the four cases separately.

Consider $S_0$.

First we evaluate the expression for the Stokes parameter
occurring in Eq.~(\ref{stokes_equal}). Dropping the angular
argument of the rotation, we get from
Eq.~(\ref{stokes_parameters})
\begin{equation}
\begin{aligned}
  S_0(RTR^{-1}\vec E)&=[RTR^{-1}\vec E]_x\overline{[RTR^{-1}\vec E]}_x\\
    &\hspace{1cm} + [RTR^{-1}\vec E]_y\overline{[RTR^{-1}\vec E]}_y\\
  &= (RTR^{-1}\vec E\,,\,RTR^{-1}\vec E),
\end{aligned}
\end{equation}
where $(\cdot,\cdot)$ denotes the standard scalar product in
$\mathbb{C}^2$. $R$ is orthogonal, thus
\begin{equation}
(RTR^{-1}\vec E\,,\,RTR^{-1}\vec E)=(TR^{-1}\vec E\,,\,TR^{-1}\vec
E),
\end{equation}
which expresses, that in a total intensity measurement, the
reflected light of a polycrystal illuminated with a single
incident beam is not distinguishable from the reflected light of a
single crystal illuminated with several beams with respective
orientations of the polarization vectors. We denote
\mbox{$c=\cos(\varphi)$}, \mbox{$s=\sin(\varphi)$} and
\begin{equation}\label{r}
  R=
  \begin{pmatrix}
    c & s\\
    -s & c\\
  \end{pmatrix}.
\end{equation}
Thus
\begin{equation}\label{t_rinv}
  TR^{-1}=
  \begin{pmatrix}
    cT_{xx}+sT_{xy} & -sT_{xx}+cT_{xy} \\
    cT_{yx}+sT_{yy} & -sT_{yx}+cT_{yy} \\
  \end{pmatrix}.
\end{equation}
Introducing $\vec E=(a,b)$, we have
\begin{equation}
\begin{split}
  [TR^{-1}\vec E]_x&\overline{[TR^{-1}\vec E]}_x
  =[(cT_{xx}+sT_{xy})a+(-sT_{xx}+cT_{xy})b]\\
   &\hspace{2cm}\times\overline{[(cT_{xx}+sT_{xy})a+(-sT_{xx}+cT_{xy})b]}\\
  =&\left(c^2T_{xx}\overline{T}_{xx}+csT_{xx}\overline{T}_{xy}+
    csT_{xy}\overline{T}_{xx}+s^2T_{xy}\overline{T}_{xy}\right)a^2\\
   &+\left(-csT_{xx}\overline{T}_{xx}+c^2T_{xx}\overline{T}_{xy}
    -s^2T_{xy}\overline{T}_{xx}+csT_{xy}\overline{T}_{xy}\right)a^2\\
   &+\left(-scT_{xx}\overline{T}_{xx}-s^2T_{xx}\overline{T}_{xy}+
    c^2T_{xy}\overline{T}_{xx}+csT_{xy}\overline{T}_{xy}\right)a^2\\
   &+\left(s^2T_{xx}\overline{T}_{xx}-scT_{xx}\overline{T}_{xy}-
    scT_{xy}\overline{T}_{xx}+c^2T_{xy}\overline{T}_{xy}\right)a^2.\\
\end{split}
\end{equation}
Thus by the first three lines of Eq.~(\ref{n_2})
\begin{equation}
\begin{aligned}
  \frac{1}{2\pi}\int_{0}^{2\pi}[T&R^{-1}(\varphi)\vec E]_x
  \overline{[TR^{-1}(\varphi)\vec E]}_x\,d\varphi\\
  &=
  \frac{1}{n}\sum_{k=1}^n [TR^{-1}({\textstyle 2\pi\frac{k}{n}})\vec E]_x
  \overline{[TR^{-1}({\textstyle 2\pi\frac{k}{n}})\vec E]}_x
\end{aligned}
\end{equation}
if and only if $n\not\in\{1,2\}$. From Eq.~(\ref{t_rinv}) we see,
that the $y$-component of the transformed vector has the same form
as the $x$-component. This completes the proof for
$S_0$.\vspace{.3cm}

Consider $S_1$ and $S_2$.

In contrast to $S_0$, both $S_1$ and $S_2$ are no scalar products.
Thus, we have to evaluate the full expression $RTR^{-1}\vec E$.
With Eq.~(\ref{r}) and Eq.~(\ref{t_rinv}) we get
\begin{equation}
\begin{aligned}
  RTR^{-1}
  =
  \Bigg(&\begin{array}{@{}c@{}}
    c^2T_{xx} + csT_{xy} + csT_{yx} + s^2T_{yy} \\
    -csT_{xx} - s^2T_{xy} + c^2T_{yx} + csT_{yy}
  \end{array} \\
  &\hspace{1cm}\begin{array}{@{}c@{}}
    -csT_{xx} + c^2T_{xy} - s^2T_{yx} + csT_{yy} \\
    s^2T_{xx} - csT_{xy} - scT_{yx} + c^2T_{yy}
  \end{array}\Bigg).
\end{aligned}
\end{equation}
We denote $\vec E=(a,b)$ as before and \mbox{$RTR^{-1}\vec E=\vec
E'=(E'_x,E'_y)$}. Then
\begin{equation}\label{eprime}
\begin{aligned}
  E'_x&=(c^2T_{xx}+csT_{xy}+csT_{yx}+s^2T_{yy})a\\
    &\hspace{1cm}+(-csT_{xx}+c^2T_{xy}-s^2T_{yx}+csT_{yy})b,\\
  E'_y&=(-csT_{xx}-s^2T_{xy}+c^2T_{yx}+csT_{yy})a\\
    &\hspace{1cm}+(s^2T_{xx}-csT_{xy}-csT_{yx}+c^2T_{yy})b.
\end{aligned}
\end{equation}
By Eq.~(\ref{stokes_parameters}), we have
\begin{align}
  S_1=&E'_x\overline{E'}_x-E'_y\overline{E'}_y
\intertext{and}
  S_2=&E'_x\overline{E'}_y+\overline{E'}_xE'_y.
\end{align}
If we evaluate these expressions by substituting
Eq.~(\ref{eprime}), every resulting term contains factors of
$\cos$ and $\sin$ of the form considered in Eq.~(\ref{n_2}) or
Eq.~(\ref{n_4}). These terms might add up to combined terms out of
the last four lines of Eq.~(\ref{n_2}) or the last three lines of
Eq.~(\ref{n_4}). To check, if we have at least one independent
term out of Eq.~(\ref{n_4}), we focus on expressions with a factor
$T_{xx}\overline{T}_{xx}$. We obtain
\begin{align}
  S_1=&T_{xx}\overline{T}_{xx}\left(c^4a^2-c^3sab+c^2s^2b^2-c^2s^2a+cs^3ab-s^4b^2\right)\notag\\
      &+\dots\notag\\
  =&T_{xx}\overline{T}_{xx}\left((c^4-c^2s^2)a^2-(s^2c^2-s^4)b^2+(cs^3-c^3s)ab\right)\notag\\
      &+\dots\\
\intertext{and}
  S_2=&2T_{xx}\overline{T}_{xx}\left(c^3sa^2+2c^2s^2ab-cs^3b^2\right)+\dots.
\end{align}
Thus, we have independent terms out of Eq.~(\ref{n_4}) which
cannot be combined to terms out of Eq.~(\ref{n_2}). This proves
the cases $S_1$ and $S_2$.\vspace{.3cm}

Last consider $S_3$.

Using the same notation as before and the results obtained in
Eq.~(\ref{eprime}), we have
\begin{align*}
  E'_x\overline{E'}_y =\big[-&c^3sT_{xx}\overline{T}_{xx}-c^2s^2T_{xx}\overline{T}_{xy}
            +c^4T_{xx}\overline{T}_{yx}+c^3sT_{xx}\overline{T}_{yy}\\
   -&c^2s^2T_{xy}\overline{T}_{xx}-cs^3T_{xy}\overline{T}_{xy}
            +c^3sT_{xy}\overline{T}_{yx}+c^2s^2T_{xy}\overline{T}_{yy}\\
   -&c^2s^2T_{yx}\overline{T}_{xx}-cs^3T_{yx}\overline{T}_{xy}
            +c^3sT_{yx}\overline{T}_{yx}+c^2s^2T_{yx}\overline{T}_{yy}\\
   -&c^2s^2T_{yy}\overline{T}_{xx}-cs^3T_{yy}\overline{T}_{xy}
            +c^3sT_{yy}\overline{T}_{yx}+c^2s^2T_{yy}\overline{T}_{yy}\big]a^2\\
  +\big[\phantom{+}&c^2s^2T_{xx}\overline{T}_{xx}-c^3sT_{xx}\overline{T}_{xy}
            -c^3sT_{xx}\overline{T}_{yx}+c^4T_{xx}\overline{T}_{yy}\\
   +&cs^3T_{xy}\overline{T}_{xx}-c^2s^2T_{xy}\overline{T}_{xy}
            -c^2s^2T_{xy}\overline{T}_{yx}+c^3sT_{xy}\overline{T}_{yy}\\
   +&cs^3T_{yx}\overline{T}_{xx}-c^2s^2T_{yx}\overline{T}_{xy}
            -c^2s^2T_{yx}\overline{T}_{yx}+c^3sT_{yx}\overline{T}_{yy}\\
   +&s^4T_{yy}\overline{T}_{xx}-cs^3T_{yy}\overline{T}_{xy}
            -cs^3T_{yy}\overline{T}_{yx}+c^2s^2T_{yy}\overline{T}_{yy}\\
   +&c^2s^2T_{xx}\overline{T}_{xx}+c^3sT_{xx}\overline{T}_{xy}
            -c^3sT_{xx}\overline{T}_{yx}-c^2s^2T_{xx}\overline{T}_{yy}\\
   -&c^3sT_{xy}\overline{T}_{xx}-c^2s^2T_{xy}\overline{T}_{xy}
            +c^4T_{xy}\overline{T}_{yx}+c^3sT_{xy}\overline{T}_{yy}\\
   +&cs^3T_{yx}\overline{T}_{xx}+s^4T_{yx}\overline{T}_{xy}
            -c^2s^2T_{yx}\overline{T}_{yx}-cs^3T_{yx}\overline{T}_{yy}\\
   -&c^2s^2T_{yy}\overline{T}_{xx}-cs^3T_{yy}\overline{T}_{xy}
            +c^3sT_{yy}\overline{T}_{yx}+c^2s^2T_{yy}\overline{T}_{yy}\big]ab\\
  +\big[-&cs^3T_{xx}\overline{T}_{xx}+c^2s^2T_{xx}\overline{T}_{xy}
            +c^2s^2T_{xx}\overline{T}_{yx}-c^3sT_{xx}\overline{T}_{yy}\\
   +&c^2s^2T_{xy}\overline{T}_{xx}-c^3sT_{xy}\overline{T}_{xy}
            -c^3sT_{xy}\overline{T}_{yx}+c^4T_{xy}\overline{T}_{yy}\\
   -&s^4T_{yx}\overline{T}_{xx}+cs^3T_{yx}\overline{T}_{xy}
            +cs^3T_{yx}\overline{T}_{yx}-c^2s^2T_{yx}\overline{T}_{yy}\\
   +&cs^3T_{yy}\overline{T}_{xx}-c^2s^2T_{yy}\overline{T}_{xy}
            -c^2s^2T_{yy}\overline{T}_{yx}+c^3sT_{yy}\overline{T}_{yy}\big]b^2.\\
\end{align*}
With Eq.~(\ref{stokes_parameters}) we obtain
\begin{align*}
  S_3=&i(E'_x\overline{E'}_y-\overline{E'}_xE'_y)\\
  =\big[&\phantom{+}(c^4+c^2s^2)T_{xx}\overline{T}_{xx}-(c^4+c^2s^2)T_{yx}\overline{T}_{xx}\\
          &+(c^3s+cs^3)T_{xx}\overline{T}_{yy}-(c^3s+cs^3)T_{yy}\overline{T}_{xx}\\
          &-(s^4+c^2s^2)T_{yy}\overline{T}_{xy}+(s^4+c^2s^2)T_{xy}\overline{T}_{yy}\\
          &+(c^3s+cs^3)T_{xy}\overline{T}_{yx}-(c^3s+cs^3)T_{yx}\overline{T}_{xy}\big]a^2\\
  +\big[&\phantom{+}2[(c^3s+cs^3)T_{yx}\overline{T}_{xx}-(c^3s+cs^3)T_{xx}\overline{T}_{yy}]\\
          &+2[(c^3s+cs^3)T_{xy}\overline{T}_{yy}-(c^3s+cs^3)T_{yy}\overline{T}_{xy}]\\
          &+(c^4+c^2s^2)T_{xx}\overline{T}_{yy}-(c^2s^2+c^4)T_{yy}\overline{T}_{xx}\\
          &+(s^4+c^2s^2)T_{yy}\overline{T}_{xx}-(c^2s^2+s^4)T_{xy}\overline{T}_{yy}\\
          &+(c^4-s^4)T_{xy}\overline{T}_{yx}+(s^4-c^4)T_{yx}\overline{T}_{xy}\big]ab\\
  +\big[&\phantom{+}(c^2s^2+s^4)T_{xx}\overline{T}_{yx}-(s^4+c^2s^2)T_{yx}\overline{T}_{xx}\\
          &+(cs^3+c^3s)T_{yx}\overline{T}_{xy}-(c^3s+cs^3)T_{xy}\overline{T}_{yx}\\
          &+(cs^3+c^3s)T_{xx}\overline{T}_{yy}-(c^3s+cs^3)T_{xx}\overline{T}_{yy}\\
          &+(c^4+c^2s^2)T_{xy}\overline{T}_{yy}-(c^2s^2+c^4)T_{yy}\overline{T}_{yx}\big]b^2.\\
\end{align*}
Indeed all terms contain a factor of $\cos$ and $\sin$ out of
those given in Eq.~(\ref{n_2}). This finishes the proof.

\bibliographystyle{apsrev}

\end{document}